\documentclass{article}%
\usepackage{amsfonts}
\usepackage{aip}
\usepackage{amsmath}
\usepackage{amssymb}
\usepackage{graphicx}%
\providecommand{\U}[1]{\protect\rule{.1in}{.1in}}
\begin{document}

\title{The Hydrodynamic Solution for Flow Profiles in a Binary Strong Electrolyte
Solution Under an External Electric Field}
\author{Byung Chan Eu\\Department of Chemistry, McGill University\\801 Sherbrooke St. West, Montreal, QC H3A 2K6\\Canada
\and Kyunil Rah\\IT and Electronic Materials R \& D\\LG Chem Research Park\\104-1 Moonji-dong, Yuseong-gu\\Daejeon 305-380, Korea}
\maketitle

\begin{abstract}
In this paper, we follow the general idea of the Onsager--Wilson theory of
strong binary electrolyte solutions and completely calculate the velocity
profile of ionic flow by first formally solving the hydrodynamic (Stokes)
equation for the ionic solutions subjected to an external electric field by a
Fourier transform method and then explicitly evaluating the formal Fourier
transform solutions as functions of spatial positions and field strength. Thus
the axial and transversal components of the velocity and the accompanying
nonequilibrium pressure are explicitly obtained. They are rare examples for
solutions of a hydrodynamic equation for flow in an external electric field.
The present results make it possible to investigate ways to overcome the
mathematical difficulty (divergence) inherent to the method of evaluating the
formal solutions that Wilson used in his dissertation on the conductance
theory (namely, the Onsager--Wilson theory) for strong binary electrolytes.
Some examples for the velocity profiles are numerically computed. They show
how ions might move in the ionic solution under an external electric field. A
possible way to get rid of the divergence causing terms for the
electrophoretic effect is examined. In the sequels, the results obtained in
this work will be applied to study ionic conductivity and related transport
processes in electrolyte solutions.

\end{abstract}

\section{\textbf{Introduction}}

\setlength{\baselineskip}{20pt}The principal initial motivation for study
underlying this work was in our desire to benefit from Onsager's theories of
conductance of electrolyte solutions which should be quite relevant to various
subjects in the fields of transport phenomena in
plasmas\cite{ting,spitzer,ichimaru}, ionic
systems\cite{mason,barthel,pombeiro,tanaka} and
semiconductors\cite{nag,landsberg,benabdallah} and micro and nano
systems\cite{ferry,karniadakis,nguyen} consisting of electrolyte solutions and
plasmas, since the conductivity and mobility of charged species subjected to
external electromagnetic fields are of great interest regardless of the scales
of the system size. At quick glance, one might think that Onsager's
theories\cite{onsager,fuoss} on the subject matter appear dated at first
glance, but the present authors believe there are some features that are still
as relevant and significant even today as before. In pursuing this line of
study we have learned that there are things of importance to improve on or
extend in his theory of electrolyte conductance and new avenues to explore as
improvement and extension unfold, as will be discussed in the course of this
line of study.

In the study of ionic conductivity, one is interested in currents carried by
ions in solution subjected to an external electric field not necessarily weak.
As the field strength increases, the conduction current has been observed to
deviate from the Ohmic law prediction. We in fact observe the well-known Wien
effect\cite{wien}, which is a non-Ohmic conductance of electrolyte solutions
subjected to high external applied electric field\cite{moore} that defy a
linear theory description. Currently, such non-Ohmic phenomena are generally
observed, for example, in semiconductors\cite{nag,landsberg,benabdallah,ferry}
and expected to be operative in micro and nanosystems because of the
necessarily large field gradients in such systems. Onsager's theory is judged
to provide valuable insights despite the differences in the sizes and states
of aggregation of the systems and hence worth a serious study.

The general ideas underlying conductance in ionic solutions, which originally
was Onsager's\cite{onsager}, consist of two effects: one is the
electrophoretic effect and the other the relaxation time effect. These can be
heuristically and qualitatively understood as follows: Let the force on ion
$j$ be
\begin{equation}
\mathbf{k}_{j}=e_{j}\mathbf{X\quad}\left(  j-1,\cdots,s\right)  \label{k1}%
\end{equation}
with $\mathbf{X}$ denoting the external electric field and $e_{j}$ the charge
of ion $j$. The ion of charge $e_{j}$ in the solution will possess an ionic
atmosphere of charge $-e_{j}$, and this atmosphere will be subjected to a
force of $-e_{j}\mathbf{X}$. This force will tend to move the atmosphere in
the direction of force $-e_{j}\mathbf{X}$, while the central ion $j$ will be
carried in the medium in a direction opposite to the motion of ion atmosphere.
The velocity of this \textit{countercurrent} may be readily calculated if it
is assumed that the entire charge $-e_{j}$ of the atmosphere is distributed in
a spherical shell of radius $\kappa^{-1}$ from the central ion and that the
motion of this sphere is governed by the Stokes law\cite{stokes,bird} holding
for the motion of a sphere in a viscous fluid. Thus, this velocity of the
countercurrent is estimated to be
\begin{equation}
\Delta\mathbf{v}_{j}=-\frac{\mathbf{k}_{j}\kappa}{6\pi\eta}=-\frac
{e_{j}\mathbf{X}\kappa}{6\pi\eta},\label{k2}%
\end{equation}
where $\Delta\mathbf{v}_{j}$ is the velocity of the interior of the shell and
$\eta_{0}$ is the viscosity of the medium. We are thus led to the result that
the medium in the interior of the shell will be travelling with this velocity,
and that the central ion will migrate against a current of this magnitude. The
deduction of this expression qualitatively elucidates the most important part
of the effect of electrophoresis. One may improve upon it by means of a
hydrodynamic method using the Navier--Stokes equations.

The second effect is: if the central ion possessed no atmosphere it would
migrate with a velocity $\mathbf{k}_{j}/\zeta_{i}$($\zeta_{i}=$ friction
constant), but owing to its atmosphere, the ion is subjected to a force,
$\mathbf{k}_{j}-\Delta\mathbf{k}_{j}$, where $\Delta\mathbf{k}_{j}$ is the
force arising from the dissymmetry of the ion atmosphere created by the
movement of the ion, and hence it will move with a velocity, relative to its
environment, of a magnitude, $\left(  \mathbf{k}_{j}-\Delta\mathbf{k}%
_{j}\right)  /\zeta_{j}$. Consequently, the net velocity $\mathbf{v}_{j}$ of
ion $j$ is given by%
\begin{equation}
\mathbf{v}_{j}=\frac{\left(  \mathbf{k}_{j}-\Delta\mathbf{k}_{j}\right)
}{\zeta_{j}}+\Delta\mathbf{v}_{j}. \label{k3}%
\end{equation}
For the case of electrical conduction, we obtain%
\begin{equation}
\mathbf{v}_{j}=\frac{\left(  e_{j}\mathbf{X}-\Delta\mathbf{k}_{j}\right)
}{\zeta_{j}}-\frac{e_{j}\mathbf{X}\kappa}{6\pi\eta}. \label{k4}%
\end{equation}
Here $\Delta\mathbf{k}_{j}\zeta_{j}^{-1}$ represents the relaxation time
effect. The aforementioned two effects underlie a qualitative explanation of
the electrophoretic and relaxation time effects in conduction.

The Onsager--Wilson (OW) theory\cite{wilson} is a theory of conductance of
strong binary electrolyte solutions in an externally applied electric field.
It formally calculates the ionic conductances by using the Onsager--Fuoss (OF)
equations\cite{fuoss} for distribution functions\ and the Poisson
equation\cite{jackson} for the modification of the field $\Delta\mathbf{k}%
_{j}$ by the presence of the ion atmosphere and, in addition to the
aforementioned two equations, the Navier--Stokes (NS) equation\cite{landau}%
---more precisely, Stokes equation---for the velocity of the medium to
calculate the electrophoretic effect. Henceforth the NS equation refers to the
Stokes equation in this article. For the NS equation the local body force
arising from the external electric field and many-body effects must be
calculated from the statistical mechanics (kinetic theory) of electrolyte
solutions. In the OW theory the local body force is calculated from the
solutions of the OF and Poisson equations. The OF equations are coupled
differential equations for pair distribution functions of ions in the
electrolyte solution, which are Fokker--Planck type equations for ions moving
like Brownian particles in the solution subjected to an electric field. It is
useful to note that, although the OF equations were originally assumed on
physical grounds they can be derived by means of the kinetic theory of dense
fluids if a Brownian motion model is assumed and the many-particle
distribution function\ is suitably approximated\cite{eu87}.

In his 1936 Yale University dissertation under the supervision of L. Onsager,
W. S. Wilson\cite{wilson} presented a formal solution for the NS equation
obtained by means of the method of Fourier transforms with the local body
force determined from the OF and Poisson equations. The formal solution
consists of sine and cosine transforms of rather complicated functions. Since
the external field is assumed directed in the positive $x$ direction of the
coordinate system and there is an axial symmetry around the $x$ axis, the
natural coordinates are cylindrical coordinates ($x,\rho,\theta$), where $x$
is the axis$\,$of the cylinder, $\rho$ is the radial coordinate perpendicular
to the cylinder axis, and $\theta$ is the azimuthal angle in the plane
perpendicular to the $x$ axis---parallel to the field direction. See Fig. 1
for the coordinates defined. The solution therefore is axially symmetric and
hence does not depend on angle $\theta$. The formal Fourier transform solution
for the NS equation obtained thereby, however, must be more explicitly
calculated as a function of $x$ and $\rho$ to make further progress in theory.
Unfortunately, the Fourier transform solution is not a kind that can be easily
evaluated in elementary functions since it consists of rather complicated
integrals. Wilson considered a special case of position coordinates for
evaluation of the Fourier transform solution.

Since the ion $j$ of ion atmosphere is the object of attention and is assumed
located at the center of the ion atmosphere, which is taken to be at the
coordinate origin, he sets $x=0$ and $\rho=0$ in the integrals making up the
solution, which then become amenable to exact and analytic evaluations. But
there is one integral among them that clearly gives rise to divergence making
the choice of $x=0$ and $\rho=0$ inappropriate.

\textit{Despite his argument to the effect that it does not contribute to the
solution because its contour integral vanishes}, it is reasonable to say in
retrospect that he simply discarded the divergent term and retained only the
convergent integrals for the calculation of electrophoresis. Despite this
troublesome aspect, the result of such a calculation was subsequently
entrenched in the literature\cite{harned} on conductance of electrolyte
solutions, especially, in connection with the Wien effect on the ionic
conductance. We find that, mathematically, this difficulty arises because the
some of the integrals involved in the Fourier transform solution are not
uniformly convergent for all values of the position coordinates. For this
reason it is not possible to set $x=0$ and $\rho=0$ in some of the integrals
before their integrations are fully performed.

We, in fact, will find that, when exactly evaluated, some of the integrals
give rise to functions diverging as $x$ and $\rho$ approach to the coordinate
origin. In other words, the solution of the NS equation for the velocity is
inherently singular at the origin of the coordinate system. It then becomes
crucially important to re-examine the OW theory to get around this difficulty
and suitably revise it before making use of it to interpret experiments on
conductance, because there arises the question of what is really meant by the
electrophoretic effect and conductance calculated therewith, given a velocity
distribution that depends on $x$ and $\rho$ and is manifestly divergent as
$x$, $\rho\rightarrow0$.

In this work, we avoid Wilson's procedure\cite{wilson} of setting $x=0$ and
$\rho=0$ within the integrals before they are evaluated, yet manage to
completely evaluate them analytically or reduce them to easily computable
quadratures. Thus we obtain relatively much simpler velocity formulas than the
Fourier transform solution that explicitly depend on $x$ and $\rho$ as well as
the field strength, for the axial and transversal velocities and the
nonequilibrium part of pressure nonlinearly depending on the external field
strength. As a matter of fact, the results obtained for the velocities and
pressure appear to be rare examples for the solutions of the NS equation of a
fluid (e.g., electrolytes) in an external electric field when the flow is
laminar that is applicable to practical experimental situations. The velocity
profiles obtained display the manner in which the ions move in the applied
electric field, and there is much insight to be gained with regard to the mode
of ionic conduction in electrolyte solutions. In the sequels to this work, we
will investigate its application to studies of ionic conductance and other
transport processes in ionic solutions.

This article is organized as follows. In Sec. II, the formal ( Fourier
transform) solutions of the OF equations for pair distribution functions and
the Poisson equations for potentials are briefly reviewed to provide the basis
for the local body force---an electric field in the present case---that is
necessary for the NS equation for velocity of the electrolyte solution in an
external electric field. The solutions of the OF and Poisson equations are
those of Wilson's in his dissertation, which have been also presented in a
recent tutorial review article\cite{euarchived} by one of the present authors.
In Sec III, the solution of such an NS equation was formally solved in the
same manner as by Wilson. We repeat his formal solution procedure, because his
work has never been published, only a brief summary of the pertinent results
having been given in the monograph of Harned and Owen\cite{harned} and in the
review article by Eckstrom and Schmeltzer\cite{eckstrom}. His
thesis\cite{wilson} itself not only skips details of the solution procedures
used, making it difficult to follow, but also contains some errors,
typographical or otherwise, in some important steps and results. Moreover, as
mentioned earlier, some of the integrals making up the formal solution diverge
at the origin of the coordinate system, causing difficulty to ascertain the
true nature and meaning of the electrophoretic effect deduced in his
dissertation. For this reason, in this work we will solve the NS equation
afresh, which by itself is quite worthwhile to learn about for its pedagogical
value for those not well versed in fluid dynamics of ionic solutions. In Sec.
IV, the formal solution consisting of a number of Fourier transforms of
complicated integrands including Bessel functions\cite{watson} of complicated
arguments will be explicitly evaluated by using the method of contour
integrations\cite{whittaker}. We thereby obtain explicit velocity profiles as
functions of coordinates and field strength for axial and transversal flows of
the electrolyte solution in a coordinate system fixed at the center ion of ion
atmosphere. Therefore the velocity profiles are of the medium with
countercharges and relative to the motion of the center ion at the coordinate
origin; this point should be remembered to avoid possible confusion about the
motions of ions in the ion atmosphere under the external electric field. In
the same section, we also calculate the nonequilibrium component of pressure
which is consistent with the velocity---the solution of the NS equation---and
depends on the field strength. In this section, we also present some numerical
examples for the axial velocity profiles that qualitatively display how ions
move under the external electric field. It reveals an interesting aspect of
ionic flow in ion atmosphere subjected to an electric field. More detailed
numerical analysis of the velocity profiles and their applications will be
made in the sequels where the theory of conductance, transport as well as
thermodynamic properties, and comparison of the theoretical result with
experiment will be discussed. In Sec. V, we examine how one might recover the
electrophoretic effect of the Onsager--Wilson theory from the result of the
present work. Sec. VI is for discussion and concluding remarks.

\section{\textbf{Pair Correlation Functions and Potentials}}

The fluid system of interest here is a binary strong electrolyte solution
subjected to an external electric field. In the OF theory of transport in
electrolyte solutions the evolution equations for ionic pair distribution
functions are assumed to obey the OF equations, which are essentially
Fokker--Planck type equations\cite{smoluchowski,chandra} in external fields of
body force. These equations are coupled to the Poisson equations for
potentials of interaction between ions. This coupled set of second-order
differential equations in fact provides an interesting and potentially
productive approach to Coulombic interaction systems subject to an external
electric field that, in our opinion, has been overlooked in the fields of
plasmas and semiconductors.

The OF equations are linear with respect to the potentials because the total
force involves the assumption that the total field due to the ions and their
atmospheres may be obtained by linear superposition of separate fields.
Onsager and Wilson\cite{wilson,euarchived} solved the coupled OF and Poisson
equations by means of Fourier transforms in a cylindrical coordinate system in
which the electric field is aligned to the axis of the cylinder. It should be
emphasized that linearizing OF equations with regard to the force does not
imply at all that the solutions of the coupled set of equations are linear
with respect to the field strength. The assumption only ensures superposition
of fields. In any case, it is the point of the Onsager--Wilson theory as far
as the OF equations for the ion pair distribution functions are concerned. The
formal solutions obtained are formally exact. They can be employed to obtain
the formal solution of the hydrodynamic equations for the velocity of the
medium. However, the formal results are impractical to use for studying
transport properties in electrolyte solutions, unless they are further
evaluated more explicitly.

Before presenting the solutions it is useful to note the symmetry properties
of the distribution functions and potentials, which are made use of to
construct the solutions. The indices $j$ and $i$ will be designated to stand
for the positive and negative ion, respectively. We denote by $\mathbf{r}$ the
relative distance vector $\mathbf{r=r}_{ij}=\mathbf{r}_{i}-\mathbf{r}%
_{j}=-\mathbf{r}_{ji}$ between ions. The ion pair distribution functions are
then denoted by $f_{ji}(\mathbf{r}_{j},\mathbf{r}_{ij})$ and $f_{ij}%
(\mathbf{r}_{i},\mathbf{r}_{ji})$, where $\mathbf{r}_{k}$ $\left(
k=i,j\right)  $ is the position vector of ion $k$.\ This implies that for the
nonuniform nonequilibrium system of interest the distribution functions are
spatially nonuniform with respect to the position of ions, e.g.,
$\mathbf{r}_{k}$ in the case of $f_{ki}(\mathbf{r}_{k},\mathbf{r}_{ik})$. The
potentials $\psi_{j}\left(  \mathbf{r}_{j},\mathbf{r}_{ij}\right)  $ and
$\psi_{i}\left(  \mathbf{r}_{i},\mathbf{r}_{ji}\right)  $ obeying the Poisson
equations also depend on the position $\mathbf{r}_{k}$ of the ion $k$ and the
relative distance vector $\mathbf{r}_{ji}=-\mathbf{r}_{ij}$ between ions $j$
and $i$. Henceforth the position vector $\mathbf{r}_{k}$ will be suppressed in
the distribution functions and potentials for the sake of notational brevity.
The Fourier transforms of solutions\cite{harned,wilson,euarchived} for the OF
equations and Poisson equations are as follows: the distribution functions are
given by%
\begin{align}
f_{ii}\left(  \mathbf{r}\right)   &  =n^{2}+\frac{2ze^{2}n^{2}}{\pi Dk_{B}%
T}\int_{0}^{\infty}d\alpha\cos\left(  \alpha x\right)  \times\nonumber\\
&  \qquad\qquad\left[  \frac{\left(  1+R\right)  }{2R^{2}}K_{0}\left(
\lambda_{1}\rho\right)  +\frac{\left(  1-R\right)  }{2R^{2}}K_{0}\left(
\lambda_{2}\rho\right)  -\frac{1-R^{2}}{R^{2}}K_{0}\left(  \lambda_{3}%
\rho\right)  \right]  ,\label{1}\\
f_{jj}\left(  \mathbf{r}\right)   &  =n^{2}-\frac{2ze^{2}n^{2}}{\pi Dk_{B}%
T}\int_{0}^{\infty}d\alpha\cos\left(  \alpha x\right)  \times\nonumber\\
&  \qquad\qquad\left[  \frac{\left(  1+R\right)  }{2R^{2}}K_{0}\left(
\lambda_{1}\rho\right)  +\frac{\left(  1-R\right)  }{2R^{2}}K_{0}\left(
\lambda_{2}\rho\right)  +\frac{1-R^{2}}{R^{2}}K_{0}\left(  \lambda_{3}%
\rho\right)  \right]  ,\label{2}%
\end{align}%
\begin{align}
f_{ij} &  =f_{ji}\left(  \pm\mathbf{r}\right)  \nonumber\\
&  =n^{2}+\frac{2n^{2}\eta^{\prime}e}{\pi D}\int_{0}^{\infty}d\alpha
\cos\left(  \alpha x\right)  \left[  \frac{\left(  1+R\right)  }{2R^{2}}%
K_{0}\left(  \lambda_{1}\rho\right)  -\frac{\left(  1-R\right)  }{2R^{2}}%
K_{0}\left(  \lambda_{2}\rho\right)  \right]  \nonumber\\
&  \qquad\;\pm\frac{2n^{2}\eta^{\prime}e\mu^{\prime}}{\pi D\kappa^{2}}\int
_{0}^{\infty}d\alpha\sin\left(  \alpha x\right)  \alpha\times\nonumber\\
&  \qquad\qquad\qquad\left[  \frac{\left(  1+R\right)  }{R^{2}}K_{0}\left(
\lambda_{1}\rho\right)  +\frac{\left(  1-R\right)  }{R^{2}}K_{0}\left(
\lambda_{2}\rho\right)  -\frac{2}{R^{2}}K_{0}\left(  \lambda_{3}\rho\right)
\right]  .\label{3}%
\end{align}
Here $f_{ii}\left(  \mathbf{r}\right)  $ and $f_{jj}\left(  \mathbf{r}\right)
$ are the pair distribution functions of two identical ions of species $i$ and
$j$ respectively, and $f_{ij}\left(  \mp\mathbf{r}\right)  $ and
$f_{ji}\left(  \pm\mathbf{r}\right)  $ are the pair distribution functions of
two different ion species $j$ and $i$ at relative distance $\mathbf{r}%
_{i}-\mathbf{r}_{j}$ and $\mathbf{r}_{j}-\mathbf{r}_{i}$, respectively. On the
other hand, the potentials of interaction---in fact, the nonequilibrium parts
thereof---are given by%
\begin{align}
\psi_{j}\left(  \pm\mathbf{r}\right)   &  =-\psi_{i}\left(  \mp\mathbf{r}%
\right)  \nonumber\\
&  =\frac{2ze}{\pi D}\int_{0}^{\infty}d\alpha\cos\left(  \alpha x\right)
\times\nonumber\\
&  \qquad\frac{1}{2R^{2}}\left[  \left(  1+R\right)  K_{0}\left(  \lambda
_{1}\rho\right)  +\left(  1-R\right)  K_{0}\left(  \lambda_{2}\rho\right)
-2\left(  1-R^{2}\right)  K_{0}\left(  \lambda_{3}\rho\right)  \right]
\nonumber\\
&  \qquad\pm\frac{2ze\mu^{\prime}}{\pi D\kappa^{2}}\int_{0}^{\infty}%
d\alpha\sin\left(  \alpha x\right)  \frac{\alpha}{R^{2}}\left[  K_{0}\left(
\lambda_{1}\rho\right)  +K_{0}\left(  \lambda_{2}\rho\right)  -2K_{0}\left(
\lambda_{3}\rho\right)  \right]  .\label{4}%
\end{align}
The potentials $\psi_{j}\left(  \pm\mathbf{r}\right)  $ and $\psi_{i}\left(
\mp\mathbf{r}\right)  $ are the nonequilibrium potentials beyond the
equilibrium potentials, which are the Debye--H\"{u}ckel potentials\cite{hill}.

In the expressions in Eqs. (\ref{1})--(\ref{4}), $x$ and $\rho$ are,
respectively, the axial and radial coordinates of the cylindrical coordinate
system; $\alpha$ is the Fourier transform variable; $K_{0}\left(  \lambda
_{l}\rho\right)  $ is the Bessel function of second kind\cite{watson} with
$\lambda_{l}$ defined as follows:%
\begin{align}
\lambda_{1}^{2} &  =\alpha^{2}+\frac{1}{2}\kappa^{2}\left(  1+R\right)
,\nonumber\\
\lambda_{2}^{2} &  =\alpha^{2}+\frac{1}{2}\kappa^{2}\left(  1-R\right)
,\label{7}\\
\lambda_{3}^{2} &  =\alpha^{2}+\frac{1}{2}\kappa^{2}.\nonumber
\end{align}
Here $R$ denotes%
\begin{equation}
R=\sqrt{1-\frac{4\mu^{\prime}{}^{2}\alpha^{2}}{\kappa^{4}}}.\label{8}%
\end{equation}
Furthermore, since the external field is assumed to be directed in the
positive $x$ direction, the functions in the present theory is axially
symmetric around the $x$ axis, and hence do not depend on the azimuthal angle
$\theta$ of the cylindrical coordinates. For the binary electrolyte under
consideration, if $z_{j}$ and $z_{i}$ are charge numbers, then $z=\left\vert
z_{j}\right\vert =\left\vert z_{i}\right\vert $; $e$ is the unit charge; $D$
is the dielectric constant of the medium; $\mu^{\prime}$ and $\eta^{\prime}$
are defined by%
\begin{equation}
\mu^{\prime}=\frac{zeX}{k_{B}T},\qquad\eta^{\prime}=\frac{ze}{k_{B}T}\label{5}%
\end{equation}
with $X$ denoting the field strength, $k_{B}$ the Boltzmann constant, $T$ the
absolute temperature; $\kappa$ is the Debye parameter of the electrolyte
solution%
\begin{equation}
\kappa=\sqrt{\frac{4\pi e^{2}n}{Dk_{B}T}\sum_{k}z_{k}^{2}c_{k}}\qquad\left(
c_{k}=\frac{n_{k}}{n}\right)  ,\label{6}%
\end{equation}
where $n$ is the density and $n_{k}$ the density of $k$. For a symmetric
binary electrolyte $\kappa$ is given by%
\[
\kappa=\sqrt{\frac{4\pi z^{2}e^{2}n}{Dk_{B}T}}\qquad\left(  z=\left\vert
z_{1}\right\vert =\left\vert z_{2}\right\vert \right)
\]

The important feature about the ion pair distribution functions and potentials
presented above is their symmetry properties with respect to interchange of
the ion positions:%
\begin{equation}
f_{ii}\left(  \mathbf{r}\right)  =f_{ii}\left(  -\mathbf{r}\right)  ,\qquad
f_{jj}\left(  \mathbf{r}\right)  =f_{jj}\left(  -\mathbf{r}\right)  ,\qquad
f_{ji}\left(  \pm\mathbf{r}\right)  =f_{ij}\left(  \mp\mathbf{r}\right)
\label{9}%
\end{equation}
and%
\begin{equation}
\psi_{j}\left(  \pm\mathbf{r}\right)  =-\psi_{i}\left(  \mp\mathbf{r}\right)
, \label{10}%
\end{equation}
and the sign ambiguity in Eqs. (\ref{3}) and (\ref{4}) refers to the ion
involved, namely, positive or negative ion. It should be remembered that the
ion positions are suppressed in the distribution functions and potentials; for
example, $f_{ji}\left(  \pm\mathbf{r}\right)  \equiv f_{ji}(\mathbf{r}%
_{j},\mathbf{r}_{ij})$ and $\psi_{j}\left(  \pm\mathbf{r}\right)
\,\equiv\,\psi_{j}\left(  \mathbf{r}_{j},\pm\mathbf{r}_{ij}\right)  $ with
$\mathbf{r=r}_{ij}$. It should be reiterated that the functions in Eqs.
(\ref{1})--(\ref{4}) do not depend on the angle variable $\theta$ owing to the
cylindrical symmetry of the system under consideration. For the derivations of
symmetry properties of the ion pair distribution functions and potentials of
interaction the reader is referred to Refs. 18, 21, and 26. Here we simply
note that the symmetry properties of Eq. (\ref{9}) and Eq. (\ref{10}) are for
interchange of ionic positions or indices for ions. Although they are not
necessary for solving the hydrodynamic equations, the Fourier transforms given
in Eqs. (\ref{1})--(\ref{4}) are explicitly evaluated by using the same method
for the velocity and pressure in Appendix A for completeness.

The Fourier transforms on the right hand sides of Eqs. (\ref{1})--(\ref{4})
represent nonequilibrium field-dependent parts of distribution functions and
potentials. Especially, Eq. (\ref{4}) must be combined with the Debye
potential if one wishes to calculate the full potential function of ion $j$.

\section{\textbf{Hydrodynamic Equation and Its Solution}}

It is well known that ions interacting through long range Coulomb potentials
produce their ion atmospheres of the mean radius given by the Debye length
$\kappa^{-1}$ centered around each of them. In an external electric field the
ion atmospheres of ions interact with the field producing \textquotedblleft
dressed\textquotedblright\ local field which in turn influences the
hydrodynamic flow and movements of ions in the solution. Such a dressed
body-force is an input when the hydrodynamics of an electrolyte solution is
sought, subject to suitable boundary conditions. Such a force must be
calculated to initiate the solution of hydrodynamic equations---namely, the NS
equation in the present case. It can be calculated by using the potential
functions given in Eq. (\ref{4}). In this study we limit the investigation to
the case of laminar flow, which allows neglecting the nonlinear inertial term
in the hydrodynamic equation. We assume that there are no body-forces other
than an electric field applied.

\subsection{\textbf{Local Electric Field}}

Assume the field is aligned along the $x$ axis. Since the charge density is
given by the Poisson equation, and the force due to the field on charge
density $\rho$ by%
\begin{equation}
F_{x}=\rho X=-\frac{DX}{4\pi}\nabla^{2}\psi\left(  \mathbf{r}\right)  ,
\label{11}%
\end{equation}
on substituting the potential function given in the previous section [see Eq.
(\ref{4})], we find the local force in the form%
\begin{align}
F_{x}  &  =-\frac{zeX}{2\pi^{2}}\int_{0}^{\infty}d\alpha\frac{1}{2R^{2}%
}\left\{  \left(  1+R\right)  \nabla^{2}\left[  \cos\left(  \alpha x\right)
K_{0}\left(  \lambda_{1}\rho\right)  \right]  \right. \nonumber\\
&  \qquad\left.  +\left(  1-R\right)  \nabla^{2}\left[  \cos\left(  \alpha
x\right)  K_{0}\left(  \lambda_{2}\rho\right)  \right]  -2\left(
1-R^{2}\right)  \nabla^{2}\left[  \cos\left(  \alpha x\right)  K_{0}\left(
\lambda_{3}\rho\right)  \right]  \right\} \nonumber\\
&  \qquad-\frac{zeX\mu^{\prime}}{2\pi^{2}\kappa^{2}}\int_{0}^{\infty}%
d\alpha\frac{\alpha}{R^{2}}\left\{  \nabla^{2}\left[  \sin\left(  \alpha
x\right)  K_{0}\left(  \lambda_{1}\rho\right)  \right]  +\nabla^{2}\left[
\sin\left(  \alpha x\right)  K_{0}\left(  \lambda_{2}\rho\right)  \right]
\right. \nonumber\\
&  \qquad\left.  -2\nabla^{2}\left[  \sin\left(  \alpha x\right)  K_{0}\left(
\lambda_{3}\rho\right)  \right]  \right\}  . \label{11a}%
\end{align}
For a binary electrolyte solution this is in fact the total local force
density. It is a \textquotedblleft dressed\textquotedblright\ force if we may
adopt a modern terminology often used in many-body physics. Since the
Laplacian operator in the cylindrical coordinates $\left(  x,\rho
,\theta\right)  $ chosen is given by
\begin{equation}
\nabla^{2}=\frac{\partial^{2}}{\partial x^{2}}+\frac{1}{\rho}\frac{\partial
}{\partial\rho}\rho\frac{\partial}{\partial\rho}+\frac{1}{\rho^{2}}%
\frac{\partial^{2}}{\partial\theta^{2}} \label{11b}%
\end{equation}
and the gradient operator by%
\begin{equation}
\mathbf{\nabla}=%
%TCIMACRO{\TeXButton{delta}{\mbox{\boldmath$\delta$}}}%
%BeginExpansion
\mbox{\boldmath$\delta$}%
%EndExpansion
_{\rho}\frac{\partial}{\partial\rho}+%
%TCIMACRO{\TeXButton{delta}{\mbox{\boldmath$\delta$}}}%
%BeginExpansion
\mbox{\boldmath$\delta$}%
%EndExpansion
_{\theta}\frac{1}{\rho}\frac{\partial}{\partial\theta}+%
%TCIMACRO{\TeXButton{delta}{\mbox{\boldmath$\delta$}}}%
%BeginExpansion
\mbox{\boldmath$\delta$}%
%EndExpansion
_{x}\frac{\partial}{\partial x}, \label{11c}%
\end{equation}
where $%
%TCIMACRO{\TeXButton{delta}{\mbox{\boldmath$\delta$}}}%
%BeginExpansion
\mbox{\boldmath$\delta$}%
%EndExpansion
_{x}$, $%
%TCIMACRO{\TeXButton{delta}{\mbox{\boldmath$\delta$}}}%
%BeginExpansion
\mbox{\boldmath$\delta$}%
%EndExpansion
_{\rho}$, and $%
%TCIMACRO{\TeXButton{delta}{\mbox{\boldmath$\delta$}}}%
%BeginExpansion
\mbox{\boldmath$\delta$}%
%EndExpansion
_{\theta}$ are unit vectors in the cylindrical coordinate system, it follows
that%
\begin{equation}
\nabla^{2}\left(
\begin{array}
[c]{c}%
\cos\left(  \alpha x\right) \\
\sin\left(  \alpha x\right)
\end{array}
\right)  K_{0}\left(  \lambda_{l}\rho\right)  =\left(
\begin{array}
[c]{c}%
\cos\left(  \alpha x\right) \\
\sin\left(  \alpha x\right)
\end{array}
\right)  \left(  \frac{1}{\rho}\frac{d}{d\rho}\rho\frac{d}{d\rho}-\alpha
^{2}\right)  K_{0}\left(  \lambda_{l}\rho\right)  . \label{12a}%
\end{equation}
However, because the Bessel function $K_{0}\left(  \lambda_{l}\rho\right)  $
obeys the differential equation\cite{watson}%
\begin{equation}
\left(  \frac{1}{\rho}\frac{d}{d\rho}\rho\frac{d}{d\rho}-\lambda_{l}%
^{2}\right)  K_{0}\left(  \lambda_{l}\rho\right)  =0, \label{12}%
\end{equation}
we obtain%
\begin{equation}
\nabla^{2}\left(
\begin{array}
[c]{c}%
\cos\left(  \alpha x\right) \\
\sin\left(  \alpha x\right)
\end{array}
\right)  K_{0}\left(  \lambda_{l}\rho\right)  =\left(  \lambda_{l}^{2}%
-\alpha^{2}\right)  \left(
\begin{array}
[c]{c}%
\cos\left(  \alpha x\right) \\
\sin\left(  \alpha x\right)
\end{array}
\right)  K_{0}\left(  \lambda_{l}\rho\right)  \label{13}%
\end{equation}
for $l=1,2,3$. Therefore the local external force is given by%
\begin{equation}
F_{x}=\overline{C}_{l}\cos\left(  \alpha x\right)  K_{0}\left(  \lambda
_{l}\rho\right)  +\overline{S}_{l}\sin\left(  \alpha x\right)  K_{0}\left(
\lambda_{l}\rho\right)  , \label{14}%
\end{equation}
where the repeated index implies a sum over the index (Einstein convention)
and symbols $\overline{C}_{l}$ and $\overline{S}_{l}$ are the abbreviations
for the following:%
\begin{equation}
\overline{C}_{l}=-\frac{zeX}{2\pi^{2}}\int_{0}^{\infty}d\alpha\left\{
\begin{array}
[c]{c}%
\frac{\left(  1+R\right)  }{2R^{2}}\left(  \lambda_{1}^{2}-\alpha^{2}\right)
\quad\quad\text{for }K_{0}(\lambda_{1}\rho)\\
\frac{\left(  1-R\right)  }{2R^{2}}\left(  \lambda_{2}^{2}-\alpha^{2}\right)
\quad\quad\text{for }K_{0}(\lambda_{2}\rho)\\
-\frac{\left(  1-R^{2}\right)  }{R^{2}}\left(  \lambda_{3}^{2}-\alpha
^{2}\right)  \quad\quad\text{for }K_{0}(\lambda_{3}\rho)
\end{array}
\right.  \label{ck}%
\end{equation}
and%
\begin{equation}
\overline{S}_{k}=-\frac{zeX}{2\pi^{2}}\frac{\mu^{\prime}}{\kappa^{2}}\int
_{0}^{\infty}d\alpha\left\{
\begin{array}
[c]{c}%
\frac{\alpha}{R^{2}}\left(  \lambda_{1}^{2}-\alpha^{2}\right)  \quad
\quad\text{for }K_{0}(\lambda_{1}\rho)\\
\frac{\alpha}{R^{2}}\left(  \lambda_{2}^{2}-\alpha^{2}\right)  \quad
\quad\text{for }K_{0}(\lambda_{2}\rho)\\
-\frac{2\alpha}{R^{2}}\left(  \lambda_{3}^{2}-\alpha^{2}\right)  \quad
\quad\text{for }K_{0}(\lambda_{3}\rho)
\end{array}
\right.  . \label{sk}%
\end{equation}
The force $F_{x}$ is an input for the NS equation of interest. The external
force $zeX$ therefore is dressed up by the extent of the integral factors in
Eqs. (\ref{ck}) and (\ref{sk}) arising from the Brownian motion of ions
interacting with the external field and among themselves through Coulomb
potentials and consequently depending on the field and density.

\subsection{\textbf{Navier--Stokes Equation}}

For a steady flow the Navier--Stokes equation takes the form
\begin{equation}
\rho\mathbf{v\cdot\nabla v}-\eta_{0}\nabla^{2}\mathbf{v-}\eta_{b}%
\mathbf{\nabla}\left(  \mathbf{\nabla\cdot v}\right)  =-\mathbf{\nabla
}p+\mathbf{F,}\label{NS}%
\end{equation}
where $\rho$ is the density, $\eta_{0}$ is the shear viscosity, $\eta_{b}$ is
the bulk viscosity, $p$ is the pressure, and $\mathbf{F}$ is the body
(external) force density. For an incompressible fluid $\mathbf{\nabla\cdot
v=\,}0$ and for a fluid undergoing laminar flow of low Reynolds number
(typically Re = $O(10^{-6})$ at the field gradient of $1$ kVolt/m in aqueous
solution) the inertial term can be neglected. Thus the Navier--Stokes
equations for velocity $\mathbf{v}$ are given by the pair of equations%
\begin{align}
-\eta_{0}\nabla^{2}\mathbf{v} &  =-\mathbf{\nabla}p+\mathbf{F,}\label{15}\\
\mathbf{\nabla\cdot v\,} &  \mathbf{=\,}0.\label{16}%
\end{align}
This set is called the Stokes equation by some authors, but we will refer to
it simply as the NS equation for an incompressible fluid in this work. Note
that the presence of an external field makes the pressure nonuniform in space.
It is interesting to note that if $curl$ of Eq. (\ref{27}) is taken, the
$\mathbf{\nabla}p$ term vanishes and Eq. (\ref{15}) takes the form%
\begin{equation}
\eta_{0}\mathbf{\nabla\times\nabla\times\nabla}\times\mathbf{v=\nabla}%
\times\mathbf{F.}\label{17a}%
\end{equation}

Since $\mathbf{\nabla\times\nabla}\times\mathbf{v=\nabla}\left(
\mathbf{\nabla\cdot v\,}\right)  -\nabla^{2}\mathbf{v}$ by vector algebra, the
two equations (\ref{15}) and (\ref{16}) may be combined into a single equation%
\begin{equation}
\eta_{0}\mathbf{\nabla\times\nabla}\times\mathbf{v=}-\mathbf{\nabla
}p+\mathbf{F.} \label{17}%
\end{equation}
This is equivalent to the Stokes equation, (\ref{15}) and (\ref{16}) for an
incompressible fluid. For the present problem $\mathbf{F=}%
%TCIMACRO{\TeXButton{delta}{\mbox{\boldmath$\delta$}}}%
%BeginExpansion
\mbox{\boldmath$\delta$}%
%EndExpansion
_{x}F_{x}$, where $%
%TCIMACRO{\TeXButton{delta}{\mbox{\boldmath$\delta$}}}%
%BeginExpansion
\mbox{\boldmath$\delta$}%
%EndExpansion
_{x}$ is the unit vector along the $x$ axis.

To solve Eq. (\ref{17}) for $\mathbf{v}$, we observe $\mathbf{\nabla\cdot
v\,}\mathbf{=\,}0$, which means that there exists an axial vector $\mathbf{A}$
such that $\mathbf{v\,}\mathbf{=\nabla\times A}$, where $\mathbf{A}$ must
depend on position vector $\mathbf{r}$ and field vector $\mathbf{X}$, both of
which are ordinary vectors, that is, polar vectors. Therefore $\mathbf{A}$
must be a vector function that must also be an axial vector, because
$\mathbf{v}$ is a polar vector. These two conditions are met simultaneously if
$\mathbf{A=\nabla\times a}$, where $\mathbf{a=a}\left(  \mathbf{r}%
,\mathbf{X}\right)  $ is a polar vector. Note in this regard that the $curl$
of an axial vector is a polar vector, and the $curl$ of an axial vector is a
polar vector. Thus we may write $\mathbf{v}$ in a general form
\begin{equation}
\mathbf{v=\nabla\times\nabla}\times\mathbf{a}+\mathbf{v}^{0}\mathbf{,}
\label{24a}%
\end{equation}
where $\mathbf{v}^{0}$ is a constant satisfying the appropriate boundary
conditions of the velocity. Note that here $\mathbf{a}$ is a polar vector.
Since $\mathbf{a\rightarrow\,}0$ and also $\mathbf{v}$ should vanish as
$\left\vert \mathbf{r}\right\vert \rightarrow\infty$, it follows
$\mathbf{v}^{0}=0$. Thus we will set $\mathbf{v}^{0}=0$ henceforth.

As the first step to formally solve Eq. (\ref{17}), substitute Eq. (\ref{24a})
with $\mathbf{v}^{0}=0$ into Eq. (\ref{17}) to obtain the equation%
\begin{equation}
\eta_{0}\mathbf{\nabla\times\nabla}\times\mathbf{\nabla\times\nabla}%
\times\mathbf{a=}-\mathbf{\nabla}p+\mathbf{F.} \label{17v}%
\end{equation}
To simplify the quadruple $curl$ term on the left hand side we observe that by
vector algebra%
\begin{align}
\mathbf{\nabla\times\nabla}\times\mathbf{a}  &  \mathbf{=\nabla}\left(
\operatorname{div}\mathbf{a}\right)  \mathbf{-\nabla}^{2}\mathbf{a,}%
\label{23a}\\
\mathbf{\nabla\times\nabla\times\nabla}\left(  \mathbf{\nabla\cdot a}\right)
&  =0. \label{23b}%
\end{align}
Then take $curl\times curl$ of Eq. (\ref{23a}) to obtain%
\begin{equation}
\mathbf{\nabla\times\nabla}\times\left(  \nabla^{2}\mathbf{a}\right)
=\mathbf{\nabla}\left(  \nabla^{2}\operatorname{div}\mathbf{a}\right)
-\mathbf{\nabla}^{2}\left(  \nabla^{2}\mathbf{a}\right)  . \label{20}%
\end{equation}
It then follows that%
\begin{align}
\mathbf{\nabla\times\nabla\times\nabla\times\nabla\times a}  &
=\mathbf{\nabla\times\nabla\times\nabla}\left(  \mathbf{\nabla\cdot a}\right)
\,\mathbf{-\,\nabla\times\nabla\times}\left(  \nabla^{2}\mathbf{a}\right)
\nonumber\\
&  =-\mathbf{\nabla\times\nabla\times}\left(  \nabla^{2}\mathbf{a}\right)  .
\label{22}%
\end{align}
Upon using Eq. (\ref{23a}) in Eq. (\ref{22}) and substituting the result into
Eq. (\ref{17v}), we obtain the equation%
\begin{equation}
\eta_{0}\nabla^{2}\nabla^{2}\mathbf{a-F=\nabla}\left(  \eta_{0}\nabla
^{2}\operatorname{div}\mathbf{a-}p\right)  , \label{17b}%
\end{equation}
which is equivalent to Eq. (\ref{17}) or the NS equation. Because the left and
right of Eq. (\ref{17b}) are of two different kinds of vectors, the equation
is satisfied if%
\begin{align}
\eta_{0}\nabla^{2}\nabla^{2}\mathbf{a}  &  =\mathbf{F,}\label{18}\\
p  &  =p_{0}+\eta_{0}\nabla^{2}\operatorname{div}\mathbf{a.}\nonumber
\end{align}
In this manner, we have deduced the equation to determine the vector
$\mathbf{a}$, namely, Eq. (\ref{18}), given the force vector $\mathbf{F}$, Eq.
(\ref{14}) in the previous subsection. Thus we have obtained the formal
solutions for $\mathbf{v}$ and $p$ satisfying the NS equation, Eq. (\ref{17}),
in terms of vector $\mathbf{a}$. The solution of the NS equation is now
reduced to that of Eq. (\ref{18}), a fourth-order differential equation, given
$\mathbf{F}$ that is provided by the OF equations and the Poisson equations.
In summary of these results, we have
\begin{align}
\mathbf{v}  &  =\mathbf{\nabla\times\nabla\times a+v}^{0}=\mathbf{\nabla
\times\nabla\times a,}\label{24}\\
p  &  =p_{0}+\eta_{0}\nabla^{2}\left(  \mathbf{\nabla\cdot a}\right)  .
\label{25}%
\end{align}
Vector $\mathbf{a}$ is determined by solving Eq. (\ref{18}) in terms of the
local force density given by Eq. (\ref{14}).

Eq. (\ref{24}) differs from Wilson's expression for $\mathbf{v}$ in two
aspects: a negative sign appears in his equation, probably a typo; and the
absence of constant term $\mathbf{v}_{0}$, which turns out equal to zero. It
turns out that the sign error in the $curl\times curl$ term on the right of
Eq. (\ref{24}) was later compensated by another sign error in the process of
determining vector $\mathbf{a}$ described in the following.

In Eq. (\ref{25}) $p_{0}$ is a homogeneous pressure uniform in space, that is,
the equilibrium pressure. This equilibrium pressure must be either supplied
phenomenologically by using thermodynamics or from the statistical mechanics
of the electrolyte solution\cite{friedman,blum}. For example, it may be
calculated from the formula%
\begin{equation}
p_{0}=nk_{B}T-\frac{2\pi}{3}\sum_{i<j}n_{i}n_{j}\int_{0}^{\infty}drr^{3}%
\frac{du_{ij}}{dr}g_{ij}^{0}\left(  r\right)  , \label{e8b}%
\end{equation}
where $u_{ij}$ is the intermolecular potential of pair $\left(  i,j\right)  $
and $g_{ij}^{0}\left(  r\right)  $ is the equilibrium pair correlation
function. The index $i$ runs over the species in the fluid, including ions.
Determination of $g_{ij}^{0}\left(  r\right)  $ should be made by following
the modern theory of equilibrium Coulomb (ionic) fluids\cite{friedman,blum}.
This homogeneous solution for $p$ is also absent in Wilson's result for $p$.

To solve Eq. (\ref{18}), substitute Eq. (\ref{14}) into the former, which then
reads%
\begin{equation}
\nabla^{2}\left(  \nabla^{2}\mathbf{a}\right)  =%
%TCIMACRO{\TeXButton{delta}{\mbox{\boldmath$\delta$}}}%
%BeginExpansion
\mbox{\boldmath$\delta$}%
%EndExpansion
_{x}C_{l}\cos\left(  \alpha x\right)  K_{0}(\lambda_{l}\rho)+%
%TCIMACRO{\TeXButton{delta}{\mbox{\boldmath$\delta$}}}%
%BeginExpansion
\mbox{\boldmath$\delta$}%
%EndExpansion
_{x}S_{l}\sin\left(  \alpha x\right)  K_{0}(\lambda_{l}\rho),\label{26}%
\end{equation}
where%
\begin{equation}
C_{l}=\frac{\overline{C}_{l}}{\eta_{0}},\qquad S_{l}=\frac{\overline{S}_{l}%
}{\eta_{0}}.\label{27}%
\end{equation}
Recalling Eq. (\ref{13}), we find
\begin{equation}
\mathbf{\nabla}^{2}\mathbf{a=}%
%TCIMACRO{\TeXButton{delta}{\mbox{\boldmath$\delta$}}}%
%BeginExpansion
\mbox{\boldmath$\delta$}%
%EndExpansion
_{x}C_{l}\frac{\cos\left(  \alpha x\right)  K_{0}(\lambda_{l}\rho)}%
{\lambda_{l}^{2}-\alpha^{2}}+%
%TCIMACRO{\TeXButton{delta}{\mbox{\boldmath$\delta$}}}%
%BeginExpansion
\mbox{\boldmath$\delta$}%
%EndExpansion
_{x}S_{l}\frac{\sin\left(  \alpha x\right)  K_{0}(\lambda_{l}\rho)}%
{\lambda_{l}^{2}-\alpha^{2}}+%
%TCIMACRO{\TeXButton{delta}{\mbox{\boldmath$\delta$}}}%
%BeginExpansion
\mbox{\boldmath$\delta$}%
%EndExpansion
_{x}A^{\ast},\label{28}%
\end{equation}
where $A^{\ast}$ is the homogeneous solution obeying the equation%
\begin{equation}
\nabla^{2}\left(  \nabla^{2}\mathbf{A}^{\ast}\right)  =0\label{29}%
\end{equation}
with $\mathbf{A}^{\ast}=%
%TCIMACRO{\TeXButton{delta}{\mbox{\boldmath$\delta$}}}%
%BeginExpansion
\mbox{\boldmath$\delta$}%
%EndExpansion
_{x}A^{\ast}$. The solution $\mathbf{A}^{\ast}$ must fit the boundary
conditions at $\rho$ infinite. Thus we choose
\begin{equation}
\mathbf{A}^{\ast}=-%
%TCIMACRO{\TeXButton{delta}{\mbox{\boldmath$\delta$}}}%
%BeginExpansion
\mbox{\boldmath$\delta$}%
%EndExpansion
_{x}C_{l}\frac{\cos\left(  \alpha x\right)  K_{0}(\alpha\rho)}{\lambda_{l}%
^{2}-\alpha^{2}}-%
%TCIMACRO{\TeXButton{delta}{\mbox{\boldmath$\delta$}}}%
%BeginExpansion
\mbox{\boldmath$\delta$}%
%EndExpansion
_{x}S_{l}\frac{\sin\left(  \alpha x\right)  K_{0}(\alpha\rho)}{\lambda_{l}%
^{2}-\alpha^{2}}.\label{30}%
\end{equation}
Therefore we obtain the equation%
\begin{equation}
\nabla^{2}\mathbf{a=}%
%TCIMACRO{\TeXButton{delta}{\mbox{\boldmath$\delta$}}}%
%BeginExpansion
\mbox{\boldmath$\delta$}%
%EndExpansion
_{x}C_{l}\frac{\cos\left(  \alpha x\right)  \left[  K_{0}(\lambda_{l}%
\rho)-K_{0}(\alpha\rho)\right]  }{\lambda_{l}^{2}-\alpha^{2}}+%
%TCIMACRO{\TeXButton{delta}{\mbox{\boldmath$\delta$}}}%
%BeginExpansion
\mbox{\boldmath$\delta$}%
%EndExpansion
_{x}S_{l}\frac{\sin\left(  \alpha x\right)  \left[  K_{0}(\lambda_{l}%
\rho)-K_{0}(\alpha\rho)\right]  }{\lambda_{l}^{2}-\alpha^{2}}.\label{31}%
\end{equation}
The solution of this inhomogeneous second-order differential equation is
sought in the form%
\begin{align}
\mathbf{a\,} &  \mathbf{=\,}%
%TCIMACRO{\TeXButton{delta}{\mbox{\boldmath$\delta$}}}%
%BeginExpansion
\mbox{\boldmath$\delta$}%
%EndExpansion
_{x}\left[  C_{l}\frac{\cos\left(  \alpha x\right)  }{\lambda_{l}^{2}%
-\alpha^{2}}+S_{l}\frac{\sin\left(  \alpha x\right)  }{\lambda_{l}^{2}%
-\alpha^{2}}\right]  \times\nonumber\\
&  \qquad\qquad\left\{  b_{1}\left[  K_{0}(\lambda_{l}\rho)-K_{0}(\alpha
\rho)\right]  +b_{2}\left[  K_{0}(\beta_{l}\rho)-K_{0}(\alpha\rho)\right]
\right\}  ,\label{h2}%
\end{align}
where $b_{1}$, $b_{2}$, and $\beta_{l}$ are constants determined as
follows---note that in Wilson's dissertation only the final results for the
coeffficients are given. On inserting this into Eq. (\ref{31}) we find%
\[
b_{1}\left(  \lambda_{l}^{2}-\alpha^{2}\right)  K_{0}(\lambda_{l}\rho
)+b_{2}\left(  \beta_{l}^{2}-\alpha^{2}\right)  K_{0}(\beta_{l}\rho
)=K_{0}(\lambda_{l}\rho)-K_{0}(\alpha\rho)
\]
or%
\begin{equation}
\left[  b_{1}\left(  \lambda_{l}^{2}-\alpha^{2}\right)  -1\right]
K_{0}(\lambda_{l}\rho)+b_{2}\left(  \beta_{l}^{2}-\alpha^{2}\right)
K_{0}(\beta_{l}\rho)+K_{0}(\alpha\rho)=0.\label{h2b}%
\end{equation}
Since $K_{0}(\lambda_{l}\rho)$, $K_{0}(\beta_{l}\rho)$, and $K_{0}(\alpha
\rho)$ are not equal to zero for all values of $\rho$, if $b_{1}$ and $b_{2}$
are chosen such that%
\begin{equation}
b_{1}=\frac{1}{\lambda_{l}^{2}-\alpha^{2}}\label{h3}%
\end{equation}
and%
\begin{equation}
\lim_{\beta_{l}\rightarrow\alpha}b_{2}\left(  \beta_{l}^{2}-\alpha^{2}\right)
K_{0}(\beta_{l}\rho)=-K_{0}(\alpha\rho),\label{h4}%
\end{equation}
then Eq. (\ref{h2}) is a solution of Eq. (\ref{31}). Eq. (\ref{h4}) implies%
\begin{equation}
b_{2}=-\frac{1}{\left(  \beta_{l}^{2}-\alpha^{2}\right)  }.\label{h5}%
\end{equation}
Finally, we obtain for the solution of Eq. (\ref{31})%
\begin{align}
\mathbf{a} &  =%
%TCIMACRO{\TeXButton{delta}{\mbox{\boldmath$\delta$}}}%
%BeginExpansion
\mbox{\boldmath$\delta$}%
%EndExpansion
_{x}\left[  C_{i}\cos\left(  \alpha x\right)  +S_{i}\sin\left(  \alpha
x\right)  \right]  \times\nonumber\\
&  \qquad\left\{  \frac{1}{\left(  \lambda_{i}^{2}-\alpha^{2}\right)  ^{2}%
}\left[  K_{0}(\lambda_{i}\rho)-K_{0}(\alpha\rho)\right]  +\frac{1}{\left(
\lambda_{i}^{2}-\alpha^{2}\right)  \alpha^{2}}\frac{\alpha\rho}{2}K_{1}%
(\alpha\rho)\right\}  .\label{32}%
\end{align}
Here again, there is a sign difference between the last term on the right of
Eq. (\ref{32}) and the corresponding term in Wilson's equation, Eq. (4.11), in
his dissertation\cite{wilson}. This particular term gives rise to the
aforementioned divergence-causing integral, but since it is essentially
ignored in his work this sign error would not affect his final result for the
electrophoretic effect.

Since the vector $\mathbf{a}$ is now obtained, it is possible to calculate the
velocity---namely, the solution of the NS equation---and pressure by using
Eqs. (\ref{24}) and (\ref{25}), respectively. For the purpose we calculate
$\operatorname{div}\mathbf{a}$, $\nabla_{x}\operatorname{div}\mathbf{a}$, and
$\nabla_{\rho}\operatorname{div}\mathbf{a}$. Since for the present problem%
\begin{equation}
\operatorname{div}\mathbf{a=}\frac{\partial a_{x}}{\partial x} \label{33}%
\end{equation}
owing to the fact that $\mathbf{F=}%
%TCIMACRO{\TeXButton{delta}{\mbox{\boldmath$\delta$}}}%
%BeginExpansion
\mbox{\boldmath$\delta$}%
%EndExpansion
_{x}F_{x}$\ and hence $a_{\rho}=a_{\theta}=0$\ identically, we find%
\begin{align}
\operatorname{div}\mathbf{a\,}  &  \mathbf{=-}\left[  C_{l}\sin\left(  \alpha
x\right)  -S_{l}\cos\left(  \alpha x\right)  \right]  \times\nonumber\\
&  \qquad\left\{  \frac{\alpha}{\left(  \lambda_{l}^{2}-\alpha^{2}\right)
^{2}}\left[  K_{0}(\lambda_{l}\rho)-K_{0}(\alpha\rho)\right]  +\frac{\rho
}{2\left(  \lambda_{l}^{2}-\alpha^{2}\right)  }K_{1}(\alpha\rho)\right\}  .
\label{34}%
\end{align}
From this follow the expressions
\begin{align}
\nabla_{x}\operatorname{div}\mathbf{a\,}  &  \mathbf{=-}\left[  C_{l}%
\cos\left(  \alpha x\right)  +S_{l}\sin\left(  \alpha x\right)  \right]
\times\nonumber\\
&  \qquad\left\{  \frac{\alpha^{2}}{\left(  \lambda_{l}^{2}-\alpha^{2}\right)
^{2}}\left[  K_{0}(\lambda_{l}\rho)-K_{0}(\alpha\rho)\right]  +\frac
{\alpha\rho}{2\left(  \lambda_{l}^{2}-\alpha^{2}\right)  }K_{1}(\alpha
\rho)\right\}  , \label{35}%
\end{align}%
\begin{align}
\nabla_{\rho}\operatorname{div}\mathbf{a\,}  &  \mathbf{=}\left[  C_{l}%
\sin\left(  \alpha x\right)  \mathbf{-}S_{l}\cos\left(  \alpha x\right)
\right]  \alpha\times\nonumber\\
&  \left\{  \frac{\lambda_{l}K_{1}(\lambda_{l}\rho)}{\left(  \lambda_{l}%
^{2}-\alpha^{2}\right)  ^{2}}-\frac{\alpha K_{1}(\alpha\rho)}{\left(
\lambda_{l}^{2}-\alpha^{2}\right)  ^{2}}+\frac{\rho}{2\left(  \lambda_{l}%
^{2}-\alpha^{2}\right)  }K_{0}\left(  \alpha\rho\right)  \right\}  ,
\label{36}%
\end{align}%
\begin{equation}
\nabla^{2}\left(  \frac{\partial a_{x}}{\partial x}\right)  \,\mathbf{=-}%
\nabla^{2}\left[  C_{l}\sin\left(  \alpha x\right)  -S_{l}\cos\left(  \alpha
x\right)  \right]  \left\{  \frac{\alpha K_{0}(\lambda_{l}\rho)}{\left(
\lambda_{l}^{2}-\alpha^{2}\right)  }-2\alpha K_{0}(\alpha\rho)\right\}  ,
\label{36p}%
\end{equation}
for which we have used the recurrence relation for the Bessel
functions\cite{watson,abramowitz}%
\begin{align}
K_{0}^{\prime}(z)  &  =-K_{1}(z),\nonumber\\
K_{1}^{\prime}\left(  z\right)   &  =-K_{0}\left(  z\right)  -\frac{1}{z}%
K_{1}\left(  z\right)  . \label{37}%
\end{align}
Here the prime denotes the derivative with respect to $z$.

\subsection{\textbf{Formal Solution for the Axial Velocity}}

It is now possible to calculate the formal solution for the axial component of
the velocity. Since%
\[
\left(  \mathbf{\nabla\times\nabla}\times\mathbf{a}\right)  _{x}%
\;\mathbf{=\nabla}_{x}\left(  \operatorname{div}\mathbf{a}\right)  -\nabla
^{2}a_{x},
\]
by using the formulas for $\mathbf{\nabla}_{x}\left(  \operatorname{div}%
\mathbf{a}\right)  $ and $\nabla^{2}a_{x}$ we obtain the formal solution for
the axial velocity component for all values of $x$ and $\rho$:%
\begin{align}
\mathbf{v}_{x}\left(  x,\rho,0\right)   &  =\mathbf{-}C_{l}\cos\left(  \alpha
x\right)  \times\nonumber\\
&  \qquad\left\{  \frac{\lambda_{l}^{2}}{\left(  \lambda_{l}^{2}-\alpha
^{2}\right)  ^{2}}\left[  K_{0}(\lambda_{l}\rho)-K_{0}(\alpha\rho)\right]
+\frac{\alpha\rho}{2\left(  \lambda_{l}^{2}-\alpha^{2}\right)  }K_{1}%
(\alpha\rho)\right\} \nonumber\\
&  \quad-S_{l}\sin\left(  \alpha x\right)  \times\nonumber\\
&  \qquad\left\{  \frac{\lambda_{l}^{2}}{\left(  \lambda_{l}^{2}-\alpha
^{2}\right)  ^{2}}\left[  K_{0}(\lambda_{l}\rho)-K_{0}(\alpha\rho)\right]
+\frac{\alpha\rho}{2\left(  \lambda_{l}^{2}-\alpha^{2}\right)  }K_{1}%
(\alpha\rho)\right\}  . \label{38}%
\end{align}
This formula does not agree with Wilson's Eq. (4.17): $K_{0}(\alpha\rho)$ is
missing in his second term, probably another typo; and the last term is
$K_{0}(\alpha\rho)$ instead of $K_{1}(\alpha\rho)$ as it is here. Since he
does not list the sine transform term (the second group of the terms in Eq.
(\ref{38})) it is not possible to compare the present formula with his formula
for the sine transform part.

On substitution of $C_{l}$ and $S_{l}$ the axial velocity formula can be
further simplified to the form%
\begin{equation}
\mathbf{v}_{x}\left(  x,\rho,0\right)  =\frac{zeX}{2\pi^{2}\eta_{0}}%
I_{c}\left(  x,\rho,0\right)  +\frac{zeX}{2\pi^{2}\eta_{0}}\left(  \frac
{\mu^{\prime}}{\kappa^{2}}\right)  I_{s}\left(  x,\rho,0\right)  ,
\label{fvax}%
\end{equation}
where%
\begin{align}
I_{c}\left(  x,\rho,0\right)   &  =\int_{0}^{\infty}d\alpha\cos\left(  \alpha
x\right)  \times\nonumber\\
&  \qquad\left\{  \frac{1}{\kappa^{2}R^{2}}\left[  \lambda_{1}^{2}%
K_{0}(\lambda_{1}\rho)+\lambda_{2}^{2}K_{0}(\lambda_{2}\rho)-2\lambda_{3}%
^{2}\left(  1-R^{2}\right)  K_{0}(\lambda_{3}\rho)\right]  \right. \nonumber\\
&  \qquad\qquad\left.  -\frac{2\lambda_{3}^{2}}{\kappa^{2}}K_{0}(\alpha
\rho)+\frac{\alpha\rho}{2}K_{1}(\alpha\rho)\right\}  \label{fvc}%
\end{align}
and%
\begin{align}
I_{s}\left(  x,\rho,0\right)   &  =\int_{0}^{\infty}d\alpha\sin\left(  \alpha
x\right)  \left\{  \frac{2\alpha}{\kappa^{2}R^{2}}\left[  \frac{\lambda
_{1}^{2}}{\left(  1+R\right)  }K_{0}(\lambda_{1}\rho)+\frac{\lambda_{2}^{2}%
}{\left(  1-R\right)  }K_{0}(\lambda_{2}\rho)\right.  \right. \nonumber\\
&  \qquad\qquad\qquad\qquad\left.  \left.  -2\lambda_{3}^{2}K_{0}(\lambda
_{3}\rho)\right]  -\frac{4\alpha^{3}}{\kappa^{2}(1-R^{2})}K_{0}(\alpha
\rho)\right\}  . \label{fvs}%
\end{align}

Wilson\cite{wilson} evaluated the integrals in $I_{c}\left(  x,\rho,0\right)
$ at $x=\rho=0$ at which point $I_{s}\left(  x,\rho,0\right)  $ is identically
equal to zero owing to the $\sin\left(  \alpha x\right)  $ factor. But the
last integrand in $I_{c}\left(  x,\rho,0\right)  $ then gives rise to
$\tfrac{1}{2}$, which yields a divergent result%
\begin{equation}
\int_{0}^{\infty}\frac{1}{2}d\alpha=\infty.\label{divi}%
\end{equation}
Wilson argued that this term does not contribute to the electrophoretic effect
because its contour integral along a semicircle in the complex plane vanishes.
However, this argument is fallacious because the integrand of this integral
does not satisfy the Jordan lemma\cite{whittaker} for the contour integral
involving a contour along an infinite semicircle.\ Moreover, even if the
contour integral along the infinite semicircle vanishes, it does not
necessarily mean the integral $\int_{0}^{\infty}\frac{1}{2}d\alpha$ vanishes.
This, on the contrary, is manifestly divergent and does not need a method of
contour integration for evaluation.

To avoid this difficulty we will evaluate the integrals in $I_{c}$ and $I_{s}$
for arbitrary values of $x$ and $\rho$. We will then apply the results to the
calculation of electrophoretic effect in the sequel; as a matter of fact, we
will have to explore a way to obtain it in a finite form.

This evaluation of $\mathbf{v}_{x}\left(  x,\rho,0\right)  $ together with
evaluations of the transversal velocity and pressure for arbitrary values of
$x$ and $\rho$ constitutes the principal contribution of this work to the
hydrodynamics of strong binary electrolyte solutions in an external electric
field. The results are the complete solution of the NS equation for binary
electrolyte solutions in an external electric field; they are not only new,
but also portend the necessity of a revision of Wilson's result for the ionic
conductivity\cite{harned,wilson}.

\subsection{\textbf{Formal Solution for the Transversal Velocity}}

By using the relation%
\begin{equation}
\left(  \mathbf{\nabla\times\nabla}\times\mathbf{a}\right)  _{\rho
}\mathbf{=\nabla}_{\rho}\left(  \operatorname{div}\mathbf{a}\right)
-\nabla^{2}\mathbf{a}_{\rho}=\mathbf{\nabla}_{\rho}\left(  \operatorname{div}%
\mathbf{a}\right)  \label{39}%
\end{equation}
in the case of $a_{\rho}=a_{\theta}=0$ we find the transversal velocity
component in the form%
\begin{equation}
\mathbf{v}_{\rho}\left(  x,\rho,0\right)  =\frac{zeX}{2\pi^{2}\eta_{0}}%
\frac{\mu^{\prime}}{\kappa^{2}}J_{c}\left(  x,\rho,0\right)  -\frac{zeX}%
{2\pi^{2}\eta_{0}}J_{s}\left(  x,\rho,0\right)  , \label{fvtr}%
\end{equation}
where%
\begin{align}
\qquad J_{c}\left(  x,\rho,0\right)   &  =\int_{0}^{\infty}d\alpha\cos\left(
\alpha x\right)  \left\{  \frac{\alpha^{2}}{\kappa^{2}R^{2}\left(
1-R^{2}\right)  }\left[  \lambda_{1}\left(  1-R\right)  K_{1}(\lambda_{1}%
\rho)\right.  \right. \nonumber\\
&  \qquad\qquad\qquad\left.  +\lambda_{2}\left(  1+R\right)  K_{1}(\lambda
_{2}\rho)-2\lambda_{3}\left(  1-R^{2}\right)  K_{1}(\lambda_{3}\rho)\right]
\nonumber\\
&  \qquad\qquad\qquad\qquad\left.  -\frac{4\alpha^{3}}{\kappa^{2}\left(
1-R^{2}\right)  }K_{1}(\alpha\rho)\right\}  ,\label{fvtc}\\
J_{s}\left(  x,\rho,0\right)   &  =\int_{0}^{\infty}d\alpha\sin\left(  \alpha
x\right)  \left\{  \frac{\alpha}{\kappa^{2}R^{2}}\left[  \lambda_{1}%
K_{1}(\lambda_{1}\rho)\right.  \right.  \times\nonumber\\
&  \qquad\qquad\qquad\qquad\left.  +\lambda_{2}K_{1}(\lambda_{2}\rho
)-2\lambda_{3}\left(  1-R^{2}\right)  K_{1}(\lambda_{3}\rho)\right]
\nonumber\\
&  \qquad\qquad\qquad\qquad\left.  -\frac{2\alpha^{2}}{\kappa^{2}}K_{1}%
(\alpha\rho)+\frac{\alpha\rho}{2}K_{0}(\alpha\rho)\right\}  . \label{fvts}%
\end{align}
The integrals in these expressions can be evaluated for all values of $x$ and
$\rho$ in a similar manner to the axial velocity, as will be shown.

\subsection{\textbf{Formal Solution for Pressure}}

Since for the present system the (nonequilibrium) pressure is given by%
\[
p=p_{0}+\eta_{0}\mathbf{\nabla}^{2}\left(  \frac{\partial a_{x}}{\partial
x}\right)  ,
\]
it is easy to calculate it from Eq. (\ref{36p}):%
\begin{align}
p-p_{0}  &  =\frac{zeX}{2\pi^{2}}\int_{0}^{\infty}d\alpha\frac{\alpha
\sin\left(  \alpha x\right)  }{2R^{2}}\left[  \left(  1+R\right)
K_{0}(\lambda_{1}\rho)\right. \nonumber\\
&  \qquad\qquad\qquad\qquad\qquad\left.  +\left(  1-R\right)  K_{0}%
(\lambda_{2}\rho)-2\left(  1-R^{2}\right)  K_{0}(\lambda_{3}\rho)\right]
\nonumber\\
&  \quad-\frac{zeX}{2\pi^{2}}\int_{0}^{\infty}d\alpha\alpha\sin\left(  \alpha
x\right)  K_{0}(\alpha\rho)\nonumber\\
&  \quad-\frac{zeX\mu^{\prime}}{2\pi^{2}\kappa^{2}}\int_{0}^{\infty}%
d\alpha\frac{\alpha^{2}\cos\left(  \alpha x\right)  }{R^{2}}\left[
K_{0}(\lambda_{1}\rho)+K_{0}(\lambda_{2}\rho)-2K_{0}(\lambda_{3}\rho)\right]
. \label{fp}%
\end{align}
The formula presented above represents a nonequilibrium part of pressure that
is consistent with the velocity components obtained as the solution of the NS
equation for a fluid in an external electric field. We have already mentioned
on how the equilibrium (homogeneous) pressure might be
calculated\cite{friedman,blum}; see Eq. (\ref{e8b}).

\section{\textbf{Evaluation of the Formal Solutions for Velocities and
Pressure}}

The formal Fourier transform solutions for the axial and transversal velocity
components and pressure, though exact, need explicit evaluation as functions
of $x$ and $\rho$ before they can be made to readily indicate their profiles
in the configuration space ($x,\rho,0$) and applied to study transport and
nonequilibrium properties of electrolyte solutions. Here we will reduce them
to either analytic forms or quadratures which can be readily evaluated by
means of simple numerical methods or approximation methods. One can, of
course, bypass this procedure and simply resort to a numerical computation
method to evaluate the formal Fourier transform solutions, but this would be
rather cumbersome and time consuming computationally, especially because most
of the integrands involved are singular at some points on the real axis and
one has to make use of the methods for singular integrals\cite{singular} to
evaluate them. The method of evaluation for the integrals discussed below
makes it unnecessary to use a direct numerical evaluation method.

It was mentioned earlier that Wilson evaluated the integrals by taking the
position variables $\left(  x,\rho\right)  $ at the coordinate origin
($x=0,\rho=0$) and that such a choice of position gives rise to a divergent
integral, Eq. (\ref{divi}). In the following we do not assume a particular set
of values for $x$ and $\rho$, but evaluate them either by means of the method
of contour integration or by conventional methods of evaluation.

It is convenient for evaluating the integrals to make use of reduced
variables. We use the following reduced variables for the purpose:%
\begin{align}
\widehat{\mathbf{v}}  &  =\left(  2\sqrt{2}\pi^{2}\eta_{0}/zeX\kappa\right)
\mathbf{v},\label{R1}\\
t  &  =\sqrt{2}\alpha/\kappa,\quad r=\rho\kappa/\sqrt{2},\quad\overline
{x}=x\kappa/\sqrt{2},\quad\xi=\mu^{\prime}/\kappa,\label{R1a}\\
\omega_{1}  &  =\frac{\sqrt{2}\lambda_{1}}{\kappa}=\sqrt{1+t^{2}+\sqrt
{1-2\xi^{2}t^{2}}},\label{R2a}\\
\omega_{2}  &  =\frac{\sqrt{2}\lambda_{2}}{\kappa}=\sqrt{1+t^{2}-\sqrt
{1-2\xi^{2}t^{2}}},\label{R2}\\
\omega_{3}  &  =\frac{\sqrt{2}\lambda_{3}}{\kappa}=\sqrt{1+t^{2}}. \label{R2b}%
\end{align}
We also use the following abbreviations:%
\begin{equation}
\overline{\omega}_{1}=\sqrt{1-y^{2}+\sqrt{1+2\xi^{2}y^{2}}},\qquad
\overline{\omega}_{3}=\sqrt{1-y^{2}},\qquad\overline{\omega}=\frac
{\sqrt{1+2\xi^{2}}}{\sqrt{2}\xi}. \label{R3}%
\end{equation}

\subsection{\textbf{Axial Velocity}}

The reduced axial velocity then takes the form%
\begin{equation}
\widehat{\mathbf{v}}_{x}\left(  x,\rho,0\right)  =\frac{1}{2}K_{c}+\frac{\xi
}{\sqrt{2}}K_{s}, \label{rv}%
\end{equation}
where%
\begin{align}
K_{c}\left(  \overline{x},r,0\right)   &  =\int_{0}^{\infty}dt\cos\left(
t\overline{x}\right)  \frac{1}{\left(  1-2\xi^{2}t^{2}\right)  }%
\times\nonumber\\
&  \qquad\qquad\left[  \omega_{1}^{2}K_{0}(\omega_{1}r)+\omega_{2}^{2}%
K_{0}(\omega_{2}r)-4\xi^{2}t^{2}\omega_{3}^{2}K_{0}(\omega_{3}r)\right]
\nonumber\\
&  \quad+2\int_{0}^{\infty}dt\cos\left(  t\overline{x}\right)  \left[
-\omega_{3}^{2}K_{0}(rt)+\frac{rt}{2}K_{1}(rt)\right]  , \label{Kc}%
\end{align}%
\begin{align}
K_{s}\left(  \overline{x},r,0\right)   &  =\int_{0}^{\infty}dt\frac
{t\sin\left(  t\overline{x}\right)  }{\left(  1-2\xi^{2}t^{2}\right)  }%
\times\nonumber\\
&  \qquad\qquad\qquad\left[  \frac{\omega_{1}^{2}K_{0}(\omega_{1}r)}{\left(
1+R\right)  }+\frac{\omega_{2}^{2}K_{0}(\omega_{2}r)}{\left(  1-R\right)
}-2\omega_{3}^{2}K_{0}(\omega_{3}r)\right] \nonumber\\
&  \quad-\frac{1}{\xi^{2}}\int_{0}^{\infty}dtt\sin\left(  t\overline
{x}\right)  K_{0}(rt). \label{Ks}%
\end{align}
It is convenient to decompose these integrals into various components for the
sake of separate evaluations as below:%
\begin{align}
K_{c}\left(  \overline{x},r,0\right)   &  =K_{1}^{c}+K_{2}^{c}-4\xi^{2}%
K_{3}^{c}-2K_{4}^{c}+rK_{5}^{c},\label{Kcd}\\
K_{s}\left(  \overline{x},r,0\right)   &  =K_{1}^{s}+K_{2}^{s}-2K_{3}%
^{s}-\frac{1}{\xi^{2}}K_{4}^{s}, \label{Ksd}%
\end{align}
where%
\begin{align}
K_{1}^{c}  &  =\int_{0}^{\infty}dt\cos\left(  t\overline{x}\right)
\frac{\omega_{1}^{2}}{\left(  1-2\xi^{2}t^{2}\right)  }K_{0}(\omega
_{1}r),\label{Kc1}\\
K_{2}^{c}  &  =\int_{0}^{\infty}dt\cos\left(  t\overline{x}\right)
\frac{\omega_{2}^{2}}{\left(  1-2\xi^{2}t^{2}\right)  }K_{0}(\omega
_{2}r),\label{Kc2}\\
K_{3}^{c}  &  =\int_{0}^{\infty}dt\cos\left(  t\overline{x}\right)
\frac{t^{2}\omega_{3}^{2}}{\left(  1-2\xi^{2}t^{2}\right)  }K_{0}(\omega
_{3}r),\label{Kc3}\\
K_{4}^{c}  &  =\int_{0}^{\infty}dt\cos\left(  t\overline{x}\right)  \omega
_{3}^{2}K_{0}(rt),\label{Kc4}\\
K_{5}^{c}  &  =\int_{0}^{\infty}dt\cos\left(  t\overline{x}\right)
tK_{1}(rt), \label{Kc5}%
\end{align}
and%
\begin{align}
K_{1}^{s}  &  =\int_{0}^{\infty}dt\sin\left(  t\overline{x}\right)
\frac{t\omega_{1}^{2}K_{0}(\omega_{1}r)}{\left(  1-2\xi^{2}t^{2}\right)
\left(  1+R\right)  },\label{Ks1}\\
K_{2}^{s}  &  =\int_{0}^{\infty}dt\sin\left(  t\overline{x}\right)
\frac{t\omega_{2}^{2}K_{0}(\omega_{2}r)}{\left(  1-2\xi^{2}t^{2}\right)
\left(  1-R\right)  },\label{Ks2}\\
K_{3}^{s}  &  =\int_{0}^{\infty}dt\sin\left(  t\overline{x}\right)
\frac{t\omega_{3}^{2}K_{0}(\omega_{3}r)}{\left(  1-2\xi^{2}t^{2}\right)
},\label{Ks3}\\
K_{4}^{s}  &  =\int_{0}^{\infty}dtt\sin\left(  t\overline{x}\right)
K_{0}(rt). \label{Ks4}%
\end{align}

Henceforth, when the reduced set of variables is used, we will omit the
overbar from $\overline{x}$ for notational brevity and understand by $x$ the
reduced variable $\overline{x}$ as defined earlier; see Eqs. (\ref{R1}%
)--(\ref{R2}).

\subsubsection{\textbf{Integrals Amenable to Conventional Methods of
Evaluation}}

We first treat $K_{4}^{c}$, $K_{5}^{c}$, and $K_{4}^{s}$, which are all
amenable to conventional methods of evaluation and give rise to elementary
functions of $x$ and $\rho$.

\paragraph{(a) $K_{4}^{c}$\newline}

The integrals listed above can be evaluated by using the integral
representation\cite{watson} of the Bessel function $K_{\nu}(rt)$ of integer
order:%
\begin{equation}
K_{\nu}(z)=\int_{0}^{\infty}dse^{-z\cosh s}\cosh\left(  \nu s\right)
\qquad\left(  \left\vert \arg z\right\vert <\frac{\pi}{2}\right)  . \label{42}%
\end{equation}
On substitution of this integral representation into integral $K_{4}^{c}$ and
interchanging the order of integration followed by change of variables we
obtain
\begin{equation}
K_{4}^{c}=\frac{\pi}{2\left(  x^{2}+r^{2}\right)  ^{1/2}}-\frac{\pi\left(
2x^{2}-r^{2}\right)  }{2\left(  r^{2}+x^{2}\right)  ^{5/2}}. \label{43}%
\end{equation}

\paragraph{(b) $K_{5}^{c}$\newline}

Upon using the integral representation of $K_{1}(rt)$ and the same procedure
as for integral $K_{4}^{c}$ to evaluate $K_{5}^{c}$, we obtain%
\begin{equation}
K_{5}^{c}=\frac{\pi r}{2\left(  x^{2}+r^{2}\right)  ^{\frac{3}{2}}}.
\label{44}%
\end{equation}

\paragraph{(c) $K_{4}^{s}$\newline}

This integral can be evaluated in the same manner as for $K_{5}^{c}$. We
obtain
\begin{equation}
K_{4}^{s}=\frac{\pi x}{2\left(  x^{2}+r^{2}\right)  ^{\frac{3}{2}}}.\label{45}%
\end{equation}
Collecting the integrals evaluated up to this point and expressing explicitly
in terms of $x$ and $r$, we find%
\begin{align}
\left(  \widehat{\mathbf{v}}_{x}\right)  _{\text{acme}} &  \equiv-K_{4}%
^{c}+\frac{r}{2}K_{5}^{c}-\frac{1}{\sqrt{2}\xi}K_{4}^{s}\nonumber\\
&  =-\frac{\pi}{2\sqrt{2}\xi}\frac{x}{\left(  x^{2}+r^{2}\right)  ^{\frac
{3}{2}}}\nonumber\\
&  \quad\,-\frac{\pi}{2\left(  x^{2}+r^{2}\right)  ^{1/2}}+\frac{\pi r^{2}%
}{4\left(  x^{2}+r^{2}\right)  ^{\frac{3}{2}}}+\frac{\pi\left(  2x^{2}%
-r^{2}\right)  }{2\left(  x^{2}+r^{2}\right)  ^{5/2}}.\label{vacm}%
\end{align}
This contribution of $\left(  \widehat{\mathbf{v}}_{x}\right)  _{\text{acme}}$
to $\widehat{\mathbf{v}}_{x}$ represents the fully deterministic part of the
hydrodynamic velocity that is not associated with the Brownian motion of
particles giving rise to the dissipative part of the local body force. In
fact, some of these terms are independent of the field, but $\left(
\widehat{\mathbf{v}}_{x}\right)  _{\text{acme}}$ is divergent at the origin.
This divergence, which is an exact result for the integrals partly making up
$\widehat{\mathbf{v}}_{x}\left(  x,r,0\right)  $, puts Wilson's result for the
electrophoresis effect in a rather vexing situation, which necessitates
careful examination of the flow profiles obtained of electrolyte solutions
before calculating the electrophoretic effect from them.

\subsubsection{\textbf{Integrals Evaluated by Means of Contour Integration
Methods}}

There are six integrals whose integrands not only are singular, but also
involve rather complicated functions and arguments for the Bessel functions
$K_{0}(\lambda_{l}\rho)$ $\left(  l=1,2,3\right)  $ in the expression for the
formal Fourier transform solution for the velocity. They defy conventional
evaluation methods, but, fortunately, methods of contour integration may be
employed for their evaluation.

To devise methods of contour integration it is necessary to learn about the
mathematical properties of the integrals involved. We list their relevant
properties below:

(1) The zeros of the arguments of the Bessel function $K_{\nu}(\lambda_{l}%
\rho)$ for $\rho\neq0$ are found to be:%
\begin{align}
\alpha &  =\pm i\sqrt{\kappa^{2}+\mu^{\prime2}}\quad\text{for }\lambda
_{1}\qquad\text{or, when reduced, }t=\pm i\sqrt{2\left(  1+\xi^{2}\right)
}\quad\text{for }\omega_{1},\label{46a}\\
\alpha &  =0\quad\text{for }\lambda_{2}\qquad\text{or }t=0\quad\text{for
}\omega_{2},\label{46b}\\
\alpha &  =\pm i\frac{\kappa}{\sqrt{2}}\quad\text{for }\lambda_{3}%
\qquad\text{or }t=\pm i\quad\text{for }\omega_{3}. \label{46c}%
\end{align}
The argument of $K_{\nu}(\lambda_{1}\rho)$ has branch points at $\alpha=\pm
i\sqrt{\kappa^{2}+\mu^{\prime2}}$, whereas the argument of $K_{\nu}%
(\lambda_{2}\rho)$ has branch points at $\alpha=0$ and $-\infty$ and the
argument of $K_{\nu}(\lambda_{3}\rho)$ has branch points at $\alpha=\pm
i\kappa/\sqrt{2}$.

Thus we may insert a branch cut on the imaginary axis of $\alpha$ plane
between $\alpha=i\sqrt{\kappa^{2}+\mu^{\prime2}}$ and $\alpha=-i\sqrt
{\kappa^{2}+\mu^{\prime2}}$ for the integral of $K_{\nu}(\lambda_{1}\rho)$,
while a branch cut may be inserted along the negative real axis for the
integral of $K_{\nu}(\lambda_{2}\rho)$, and on the imaginary axis between
$\alpha=i\kappa/\sqrt{2}$ and $\alpha=-i\kappa/\sqrt{2}$ for the integral of
$K_{\nu}(\lambda_{3}\rho)$, respectively. See Figs 2--4 below.

(2) We recall that Bessel function $K_{\nu}(z)$ of complex variable $z$ is
regular in $z$ plane cut along negative real axis\cite{watson,abramowitz}.
That is, the Bessel function is a multi-valued function in the cut plane. In
the present case, $K_{0}(\lambda_{1}\rho)$ changes discontinuously as the
branch cut $\left[  -i\sqrt{\kappa^{2}+\mu^{\prime2}},i\sqrt{\kappa^{2}%
+\mu^{\prime2}}\right]  $ is crossed, whereas $K_{\nu}(\lambda_{2}\rho)$
changes discontinuously as the negative real axis is crossed, and $K_{\nu
}(\lambda_{3}\rho)$ changes discontinuously as the branch cut $\left[
-i\frac{1}{\sqrt{2}}\kappa,i\frac{1}{\sqrt{2}}\kappa\right]  $ is crossed on
the imaginary axis. Note that the Bessel functions $K_{\nu}(\lambda_{2}\rho)$
and $K_{\nu}\left(  \alpha\rho\right)  $ are defined in $\alpha$ plane cut
along the negative real axis.

(3) We also observe that all the integrands of the singular integrals in
$I_{c}$ and $I_{s}$ in Eqs. (\ref{fvc}) and (\ref{fvs}) are even with respect
to $\alpha$.

(4) Moreover, for $0<\arg\alpha<\pi$ we find%
\begin{equation}
-\left(  \frac{\pi}{2}-\delta\right)  <\arg\lambda_{l}<\left(  \frac{\pi}%
{2}-\delta\right)  \quad\left(  \frac{\pi}{2}>\delta>0;\;l=1,2,3\right)  .
\label{47}%
\end{equation}
Therefore in the upper half plane of complex $\alpha$
\begin{equation}
\lim_{\left\vert \alpha\right\vert \rightarrow\infty}K_{\nu}(\lambda_{l}%
\rho)=\lim_{\left\vert \alpha\right\vert \rightarrow\infty}\sqrt{\frac{\pi
}{2\lambda_{l}\rho}}e^{-\lambda_{l}\rho}\rightarrow0. \label{48}%
\end{equation}
By this, if contours for the singular integrals are taken along an infinite
semicircle in the upper $\alpha$ plane, Jordan's lemma\cite{whittaker} for
contour integration along a circle of infinite radius is assuredly satisfied.

(5) Lastly, all the integrands in Eqs. (\ref{Kc1})--(\ref{Kc3}) and
(\ref{Ks1})--(\ref{Ks3}) have simple poles at
\begin{equation}
\alpha=\pm\frac{\kappa^{2}}{2\mu^{\prime}}\quad\text{or }t=\pm\frac{1}%
{\sqrt{2}\xi}. \label{49}%
\end{equation}

There is also a branch cut between $t=-\frac{1}{\sqrt{2}\xi}$ and $t=\frac
{1}{\sqrt{2}\xi}$, but this particular branch cut associated with
$\sqrt{1-2\xi^{2}t^{2}}$ does not play a role in the contour integrals
considered in the present work, because the real axis is not crossed by the
contours in performing integrations.

All these properties (1)--(5) together suggest it is possible to evaluate the
integrals by using methods of contour integration\cite{whittaker} along the
closed contours of infinite semicircle as depicted in Figs. 2--4. This is
indeed a fortunate combination of properties of the integrands involved. We
will indicate which contour applies to which integral at appropriate points in
the discussion. Since methods of integration will be similar for the integrals
involved in $K_{l}^{c}$ and $K_{l}^{s}$ $\left(  l=1,2,3\right)  $ we will
illustrate them with the examples of integrals in $K_{1}^{c}$ and $K_{1}^{s}$
in Appendix A. The results for the rest of integrals will be simply presented
with appropriate comments on the use of contours in Figs. 2--4 and useful
points if needed. And then by combining the results for the integrals the
final formula for the axial and transversal velocities and pressure will be
presented. Incidentally, the same contours can be used for evaluating the
Fourier transforms for the pair distributions and potentials in Eq.
(\ref{1})--(\ref{4}) as shown for completeness in Appendix A.

\paragraph{(d) $K_{1}^{c}$ and $K_{1}^{s}$\textit{\newline}}

As prototypes of contour integrals appearing in present work, integrals
$K_{1}^{c}$ and $K_{1}^{s}$ are explicitly evaluated in Appendix A. Integrals
$K_{1}^{c}$ and $K_{1}^{s}$ both have simple poles at $t=\pm\left(  \sqrt
{2}\xi\right)  ^{-1}$. There is a branch cut along the imaginary axis between
$t=-i\sqrt{2\left(  1+\xi^{2}\right)  }$ and $+i\sqrt{2\left(  1+\xi
^{2}\right)  }$ and also a branch cut on the real axis between $t=-1/\sqrt
{2}\xi$ and $t=+1/\sqrt{2}\xi$, but the latter branch cut plays no role in
integration since the path of integration in all the contours does not cross
it. For this reason the latter branch cut is not shown in Figs. 2--4. For
evaluation of both $K_{1}^{c}$ and $K_{1}^{s}$ the contour in Fig. 2 is used.
The results are as follows:%
\begin{align}
K_{1}^{c}  &  =-\frac{\sqrt{2}\pi\left(  1+2\xi^{2}\right)  }{8\xi^{3}}%
\sin\left(  \frac{x}{\sqrt{2}\xi}\right)  K_{0}(\overline{\omega}r)\nonumber\\
&  \quad-\frac{\pi}{2}\int_{0}^{\sqrt{2\left(  1+\xi^{2}\right)  }}%
dy\frac{e^{-xy}\left(  1-y^{2}+\sqrt{1+2\xi^{2}y^{2}}\right)  }{1+2\xi
^{2}y^{2}}I_{0}(\overline{\omega}_{1}r), \label{50c}%
\end{align}
and
\begin{align}
K_{1}^{s}  &  =\frac{\pi\left(  1+2\xi^{2}\right)  }{8\xi^{4}}\cos\left(
xt\right)  K_{0}(\overline{\omega}r)\nonumber\\
&  \quad-\frac{\pi}{2}\int_{0}^{\sqrt{2\left(  1+\xi^{2}\right)  }}%
dy\frac{e^{-xy}y\left(  1-y^{2}+\sqrt{1+2\xi^{2}y^{2}}\right)  }{\left(
1+2\xi^{2}y^{2}\right)  \left(  1+\sqrt{1+2\xi^{2}y^{2}}\right)  }%
I_{0}(\overline{\omega}_{1}r). \label{50s}%
\end{align}
The integrals in these formulas arise from integration along the branch cut.
See Appendix A for the details of calculation. Therefore%
\begin{align}
\frac{1}{2}K_{1}^{c}+\frac{\xi}{\sqrt{2}}K_{1}^{s}  &  =-\frac{\pi\left(
1+2\xi^{2}\right)  }{8\sqrt{2}\xi^{3}}\left[  \sin\left(  \frac{x}{\sqrt{2}%
\xi}\right)  -\cos\left(  xt\right)  K_{0}(\overline{\omega}r)\right]
K_{0}(\overline{\omega}r)\nonumber\\
&  \quad-\frac{\pi}{4}\int_{0}^{\sqrt{2\left(  1+\xi^{2}\right)  }}%
dy\frac{e^{-xy}\left(  1-y^{2}+\sqrt{1+2\xi^{2}y^{2}}\right)  }{1+2\xi
^{2}y^{2}}\times\nonumber\\
&  \qquad\qquad\qquad\left[  1-\frac{\sqrt{2}\xi y}{\left(  1+\sqrt{1+2\xi
^{2}y^{2}}\right)  }\right]  I_{0}(\overline{\omega}_{1}r). \label{51c}%
\end{align}

\paragraph{(e) $K_{2}^{c}$\textbf{ and }$K_{2}^{s}$\textit{\newline}}

For evaluation of these integrals the contour in Fig. 3 is used since the
integrand does not have a branch cut on the imaginary axis, but there is
contributions from the residues. They give rise to the following results:%
\begin{align}
K_{2}^{c}  &  =-\frac{\sqrt{2}\pi\left(  1+2\xi^{2}\right)  }{8\xi^{3}}%
\sin\left(  \frac{x}{\sqrt{2}\xi}\right)  K_{0}(\overline{\omega
}r),\label{K2c}\\
K_{2}^{s}  &  =\frac{\pi\left(  1+2\xi^{2}\right)  }{8\xi^{4}}\cos\left(
\frac{x}{\sqrt{2}\xi}\right)  K_{0}(\overline{\omega}r). \label{K2s}%
\end{align}
Therefore we obtain%
\begin{equation}
\frac{1}{2}K_{2}^{c}+\frac{\xi}{\sqrt{2}}K_{2}^{s}=-\frac{\pi\left(
1+2\xi^{2}\right)  }{8\sqrt{2}\xi^{3}}\left[  \sin\left(  \frac{x}{\sqrt{2}%
\xi}\right)  -\cos\left(  \frac{x}{\sqrt{2}\xi}\right)  \right]
K_{0}(\overline{\omega}r). \label{K2cs}%
\end{equation}

\paragraph{(f) $K_{3}^{c}$\textbf{ and }$K_{3}^{s}$\textit{\newline}}

The integrands of these integrals involve a branch cut along the imaginary
axis from $z=-i$ to $+i$. Therefore the appropriate contour to use is
$\mathcal{C}_{3}$ depicted in Fig. 4. The results of their evaluation are as
follows:%
\begin{align}
K_{3}^{c}  &  =-\frac{\sqrt{2}\pi}{16\xi^{5}}\left(  1+2\xi^{2}\right)
K_{0}(\overline{\omega}r)\sin\left(  \frac{x}{\sqrt{2}\xi}\right) \nonumber\\
&  \qquad+\frac{\pi}{2}\int_{0}^{1}dy\frac{e^{-xy}y^{2}\left(  1-y^{2}\right)
}{1+2\xi^{2}y^{2}}I_{0}(\overline{\omega}_{3}r),\label{K3c}\\
K_{3}^{s}  &  =\frac{\pi}{8\xi^{4}}\left(  1+2\xi^{2}\right)  K_{0}%
(\overline{\omega}r)\cos\left(  \frac{x}{\sqrt{2}\xi}\right) \nonumber\\
&  \qquad-\frac{\pi}{2}\int_{0}^{1}dy\frac{e^{-xy}y\left(  1-y^{2}\right)
}{1+2\xi^{2}y^{2}}I_{0}(\overline{\omega}_{3}r). \label{K3s}%
\end{align}
Therefore we find%
\begin{align}
-2\xi^{2}K_{3}^{c}-\sqrt{2}\xi K_{3}^{s}  &  =\frac{\pi}{4\sqrt{2}\xi^{3}%
}\left(  1+2\xi^{2}\right)  K_{0}(\overline{\omega}r)\left[  \sin\left(
\frac{x}{\sqrt{2}\xi}\right)  -\cos\left(  \frac{x}{\sqrt{2}\xi}\right)
\right] \nonumber\\
&  \quad-\frac{\pi}{2}\int_{0}^{1}dy\frac{e^{-xy}\sqrt{2}\xi y\left(
1-y^{2}\right)  \left(  \sqrt{2}\xi y-1\right)  }{1+2\xi^{2}y^{2}}%
I_{0}(\overline{\omega}_{3}r). \label{K3cs}%
\end{align}

\paragraph{Summary for the Reduced Axial Velocity\textit{\newline}}

Collecting the results presented earlier, we obtain the reduced axial velocity%
\begin{align}
\widehat{\mathbf{v}}_{x}\left(  x,\rho,0\right)   &  =-\frac{\pi}{2\sqrt{2}%
\xi}\frac{x}{\left(  x^{2}+r^{2}\right)  ^{3/2}}\nonumber\\
&  \qquad-\frac{\pi}{2\left(  x^{2}+r^{2}\right)  ^{1/2}}+\frac{\pi r^{2}%
}{4\left(  x^{2}+r^{2}\right)  ^{3/2}}+\frac{\pi\left(  2x^{2}-r^{2}\right)
}{2\left(  x^{2}+r^{2}\right)  ^{5/2}}\nonumber\\
&  \qquad-\frac{\pi}{4}\int_{0}^{\sqrt{2\left(  1+\xi^{2}\right)  }}%
dy\frac{e^{-xy}\left(  1-y^{2}+\sqrt{1+2\xi^{2}y^{2}}\right)  }{1+2\xi
^{2}y^{2}}\times\nonumber\\
&  \qquad\qquad\qquad\qquad\quad\left[  1-\frac{\sqrt{2}\xi y}{\left(
1+\sqrt{1+2\xi^{2}y^{2}}\right)  }\right]  I_{0}(\overline{\omega}%
_{1}r)\nonumber\\
&  \qquad-\frac{\pi}{2}\int_{0}^{1}dy\frac{e^{-xy}\sqrt{2}\xi y\left(
1-y^{2}\right)  }{1+2\xi^{2}y^{2}}\left[  \sqrt{2}\xi y-1\right]
I_{0}(\overline{\omega}_{3}r). \label{rvax}%
\end{align}
We note that the terms made up of trigonometric functions in Eqs. (\ref{51c}),
(\ref{K2cs}), and (\ref{K3cs}) cancel each other out. This velocity formula
(\ref{rvax}) is the velocity profile of the countercurrent of the ion
atmosphere in the coordinate system fixed at the center ion of the ion
atmosphere pulled by the external electric field. The first four terms on the
right represent a \textquotedblleft deterministic\textquotedblright\ part of
the velocity $\left(  \widehat{\mathbf{v}}_{x}\right)  _{\text{acme}}$ and the
two integrals involving the Bessel functions $I_{0}(\overline{\omega}_{1}r)$
and $I_{0}(\overline{\omega}_{3}r)$ stem from the Brownian motion part of the
local force---namely, the dressed-up part of the force arising from the
interaction of the ion atmosphere and the external electric field.

\subsection{\textbf{Transversal Velocity}}

On reducing the integrals, we obtain the reduced transversal velocity as%
\begin{align}
\widehat{\mathbf{v}}_{\rho}\left(  x,\rho,0\right)   &  =\frac{\xi}{2\sqrt{2}%
}\left(  J_{1}^{c}+J_{2}^{c}-2J_{3}^{c}\right)  -\frac{1}{\sqrt{2}\xi}%
J_{4}^{c}\nonumber\\
&  -\frac{1}{2}\left(  J_{1}^{s}+J_{2}^{s}-4\xi^{2}J_{3}^{s}\right)
+J_{4}^{s}-\frac{1}{2}rJ_{5}^{s}, \label{rvr}%
\end{align}
where the component integrals are defined by
\begin{align}
J_{1}^{c}  &  =\int_{0}^{\infty}dt\cos\left(  t\overline{x}\right)
\frac{t^{2}\omega_{1}}{\left(  1-2\xi^{2}t^{2}\right)  \left(  1+R\right)
}K_{1}(\omega_{1}r),\label{J1c}\\
J_{2}^{c}  &  =\int_{0}^{\infty}dt\cos\left(  t\overline{x}\right)
\frac{t^{2}\omega_{1}}{\left(  1-2\xi^{2}t^{2}\right)  \left(  1-R\right)
}K_{1}(\omega_{2}r),\label{J2c}\\
J_{3}^{c}  &  =\int_{0}^{\infty}dt\cos\left(  t\overline{x}\right)
\frac{t^{2}}{\left(  1-2\xi^{2}t^{2}\right)  }\omega_{3}K_{1}(\omega
_{3}r),\label{J3c}\\
J_{4}^{c}  &  =\int_{0}^{\infty}dt\cos\left(  t\overline{x}\right)
tK_{1}(tr), \label{J4c}%
\end{align}
and%
\begin{align}
J_{1}^{s}  &  =\int_{0}^{\infty}dt\sin\left(  t\overline{x}\right)
\frac{t\omega_{1}}{\left(  1-2\xi^{2}t^{2}\right)  }K_{1}(\omega
_{1}r),\label{J1s}\\
J_{2}^{s}  &  =\int_{0}^{\infty}dt\sin\left(  t\overline{x}\right)
\frac{t\omega_{2}}{\left(  1-2\xi^{2}t^{2}\right)  }K_{1}(\omega
_{2}r),\label{J2s}\\
J_{3}^{s}  &  =\int_{0}^{\infty}dt\sin\left(  t\overline{x}\right)
\frac{t^{3}\omega_{3}}{\left(  1-2\xi^{2}t^{2}\right)  }K_{1}(\omega
_{3}r),\label{J3s}\\
J_{4}^{s}  &  =\int_{0}^{\infty}dt\sin\left(  t\overline{x}\right)  t^{2}%
K_{1}(rt),\label{J4s}\\
J_{5}^{s}  &  =\int_{0}^{\infty}dt\sin\left(  t\overline{x}\right)
tK_{0}(rt). \label{J5s}%
\end{align}
These integrals can be evaluated in the same manner as for the axial velocity
despite the fact that most of them are given in terms of Bessel functions one
order higher than $K_{0}\left(  z\right)  $ appearing in the axial velocity
integrals, namely, $K_{1}(z)$. The integrals $J_{l}^{c}$ and $J_{l}^{s}$ for
$l=1,2,3$ are evaluated by using the contours in Figs. 2, 3, and 4,
respectively, and integrals $J_{4}^{c}$, $J_{4}^{s}$, and $J_{5}^{s}$ are
evaluated by the conventional method using the integral representation of
$K_{1}(z)$. The results are as follows:%
\begin{align}
J_{1}^{c}  &  =-\frac{\pi\left(  2\xi^{2}+1\right)  }{8\sqrt{2}\xi^{5}}%
K_{1}(\overline{\omega}r)\sin\left(  \frac{x}{\sqrt{2}\xi}\right) \nonumber\\
&  \qquad+\frac{\pi}{2}\int_{0}^{\sqrt{2\left(  1+\xi^{2}\right)  }}%
dy\frac{e^{-xy}y^{2}\overline{\omega}_{1}I_{1}(\overline{\omega}_{1}%
r)}{\left(  1+2\xi^{2}y^{2}\right)  \left(  1+\sqrt{1+2\xi^{2}y^{2}}\right)
},\label{J1ce}\\
J_{2}^{c}  &  =-\frac{\pi}{8\sqrt{2}\xi^{5}}\left(  2\xi^{2}+1\right)
K_{1}(\overline{\omega}r)\sin\left(  \frac{x}{\sqrt{2}\xi}\right)
,\label{J2ce}\\
J_{3}^{c}  &  =-\frac{\pi}{8\sqrt{2}\xi^{5}}\left(  2\xi^{2}+1\right)
K_{1}(\overline{\omega}r)\sin\left(  \frac{x}{\sqrt{2}\xi}\right) \nonumber\\
&  \qquad+\frac{\pi}{2}\int_{0}^{1}dye^{-xy}\frac{y^{2}\left(  1-y^{2}\right)
}{\left(  1+2\xi^{2}y^{2}\right)  }I_{1}(\overline{\omega}_{3}r),
\label{J3ce}\\
J_{4}^{c}  &  =\frac{\pi}{2r^{2}\left(  u^{2}+1\right)  ^{\frac{3}{2}}}%
=\frac{\pi r}{2\left(  x^{2}+r^{2}\right)  ^{\frac{3}{2}}}. \label{J4ce}%
\end{align}
Similarly, we obtain%
\begin{align}
J_{1}^{s}  &  =\frac{\pi}{8\xi^{4}}\left(  2\xi^{2}+1\right)  K_{1}%
(\overline{\omega}r)\cos\left(  \frac{x}{\sqrt{2}\xi}\right) \nonumber\\
&  \qquad-\frac{1}{2}\pi\int_{0}^{\sqrt{2\left(  1+\xi^{2}\right)  }}%
dy\frac{e^{-xy}y\overline{\omega}_{1}}{1+2\xi^{2}y^{2}}I_{1}(\overline{\omega
}_{1}r),\label{J1se}\\
J_{2}^{s}  &  =\frac{\pi}{8\xi^{4}}\left(  2\xi^{2}+1\right)  K_{1}%
(\overline{\omega}r)\cos\left(  \frac{x}{\sqrt{2}\xi}\right)  ,\label{J2se}\\
J_{3}^{s}  &  =\frac{\pi}{16\xi^{6}}\left(  2\xi^{2}+1\right)  K_{1}%
(\overline{\omega}r)\cos\left(  \frac{x}{\sqrt{2}\xi}\right) \nonumber\\
&  \qquad+\frac{1}{2}\pi\int_{0}^{1}dye^{-xy}\frac{y^{3}\overline{\omega}_{3}%
}{1+2\xi^{2}y^{2}}I_{1}(\overline{\omega}_{3}r),\label{J3se}\\
J_{4}^{s}  &  =-\frac{3\pi x\left(  x^{4}-x^{2}r^{2}-r^{4}\right)  }%
{2r^{3}\left(  x^{2}+r^{2}\right)  ^{\frac{5}{2}}},\label{J4se}\\
J_{5}^{s}  &  =\frac{\pi x}{2\left(  x^{2}+r^{2}\right)  ^{3/2}}. \label{J5se}%
\end{align}
In summary, we obtain the reduced transversal velocity%
\begin{align}
\widehat{\mathbf{v}}_{\rho}\left(  x,\rho,0\right)   &  =-\frac{\pi r}%
{2\sqrt{2}\xi\left(  x^{2}+r^{2}\right)  ^{\frac{3}{2}}}-\frac{\pi x}{2\left(
x^{2}+r^{2}\right)  ^{3/2}}\left[  \frac{r}{2}+\frac{3\left(  x^{4}-x^{2}%
r^{2}-r^{4}\right)  }{r^{3}\left(  x^{2}+r^{2}\right)  }\right] \nonumber\\
&  +\frac{\pi}{4}\int_{0}^{\sqrt{2\left(  1+\xi^{2}\right)  }}dy\frac
{e^{-xy}y\overline{\omega}_{1}}{\left(  1+2\xi^{2}y^{2}\right)  }\left(
1+\frac{\sqrt{2}\xi y}{2\left(  1+\sqrt{1+2\xi^{2}y^{2}}\right)  }\right)
I_{1}(\overline{\omega}_{1}r)\nonumber\\
&  +\pi\xi^{2}\int_{0}^{1}dye^{-xy}\frac{y^{2}\left(  y-\sqrt{2}\xi\right)
\left(  1-y^{2}\right)  }{1+2\xi^{2}y^{2}}I_{1}(\overline{\omega}_{3}r).
\label{rvrfin}%
\end{align}
As is the case for the axial velocity, the trigonometric function terms in
Eqs. (\ref{J1ce})--(\ref{J3ce}) and Eqs. (\ref{J1se})--(\ref{J3se}) cancel in
the formula for the transversal velocity formula. As is evident from this
expression, the transversal velocity also diverges at the origin similarly to
the axial velocity does.

\subsection{\textbf{Pressure}}

The expression for nonequilibrium pressure $\Delta p=p-p_{0}$ also can be
decomposed into various reduced integrals for the purpose of evaluation:%
\begin{equation}
\Delta\widehat{p}=\frac{1}{2}\left(  P_{1}^{s}+P_{2}^{s}\right)  -2\xi
^{2}P_{3}^{s}-P_{4}^{s}-\frac{\xi}{\sqrt{2}}\left(  P_{1}^{c}+P_{2}%
^{c}\right)  +\sqrt{2}\xi P_{3}^{c},\label{delp}%
\end{equation}
where $\Delta\widehat{p}$ is the reduced pressure%
\begin{equation}
\Delta\widehat{p}=\Delta p\left(  \frac{zeX\kappa^{2}}{4\pi^{2}}\right)
^{-1}\label{rdp}%
\end{equation}
and%
\begin{align}
P_{1}^{c} &  =\int_{0}^{\infty}dt\cos\left(  tx\right)  \frac{t^{2}}{\left(
1-2\xi^{2}t^{2}\right)  }K_{0}(\omega_{1}r),\label{P1c}\\
P_{2}^{c} &  =\int_{0}^{\infty}dt\cos\left(  tx\right)  \frac{t^{2}}{\left(
1-2\xi^{2}t^{2}\right)  }K_{0}(\omega_{2}r),\label{P2c}\\
P_{3}^{c} &  =\int_{0}^{\infty}dt\cos\left(  tx\right)  \frac{t^{2}}{\left(
1-2\xi^{2}t^{2}\right)  }K_{0}(\omega_{3}r),\label{P3c}%
\end{align}%
\begin{align}
P_{1}^{s} &  =\int_{0}^{\infty}dt\sin\left(  tx\right)  \frac{t\left(
1+R\right)  }{\left(  1-2\xi^{2}t^{2}\right)  }K_{0}(\omega_{1}r),\label{P1s}%
\\
P_{2}^{s} &  =\int_{0}^{\infty}dt\sin\left(  tx\right)  \frac{t\left(
1-R\right)  }{\left(  1-2\xi^{2}t^{2}\right)  }K_{0}(\omega_{2}r),\label{P2s}%
\\
P_{3}^{s} &  =\int_{0}^{\infty}dt\sin\left(  tx\right)  \frac{t^{3}}{\left(
1-2\xi^{2}t^{2}\right)  }K_{0}(\omega_{3}r),\label{P3s}\\
P_{4}^{s} &  =\int_{0}^{\infty}dtt\sin\left(  tx\right)  K_{0}(rt).\label{P4s}%
\end{align}
These integrals are evaluated similarly to the velocities presented earlier:%
\begin{align}
P_{1}^{c} &  =-\frac{\pi}{4\sqrt{2}\xi^{3}}K_{0}(\overline{\omega}%
r)\sin\left(  \frac{x}{\sqrt{2}\xi}\right)  \nonumber\\
&  \qquad+\frac{\pi}{2}\int_{0}^{\sqrt{2\left(  1+\xi^{2}\right)  }}%
dye^{-xy}\frac{y^{2}}{1+2\xi^{2}y^{2}}I_{0}(\overline{\omega}_{1}%
r),\label{P1ce}\\
P_{2}^{c} &  =-\frac{\pi}{4\sqrt{2}\xi^{3}}K_{0}(\overline{\omega}%
r)\sin\left(  \frac{x}{\sqrt{2}\xi}\right)  ,\label{P2ce}\\
P_{3}^{c} &  =-\frac{\pi}{4\sqrt{2}\xi^{3}}K_{0}(\overline{\omega}%
r)\sin\left(  \frac{x}{\sqrt{2}\xi}\right)  \nonumber\\
&  \qquad+\frac{\pi}{2}\int_{0}^{1}dye^{-xy}\frac{y^{2}}{1+2\xi^{2}y^{2}}%
I_{0}(\overline{\omega}_{3}r),\label{P3ce}%
\end{align}%
\begin{align}
P_{1}^{s} &  =\frac{\pi}{4\xi^{2}}K_{0}(\overline{\omega}r)\cos\left(
\frac{x}{\sqrt{2}\xi}\right)  \nonumber\\
&  \quad-\frac{\pi}{2}\int_{0}^{\sqrt{2\left(  1+\xi^{2}\right)  }}%
dye^{-xy}\frac{y\left(  1+\sqrt{1+2\xi^{2}y^{2}}\right)  }{1+2\xi^{2}y^{2}%
}I_{0}(\overline{\omega}_{1}r),\label{P1se}\\
P_{2}^{s} &  =\frac{\pi}{4\xi^{2}}K_{0}(\overline{\omega}r)\cos\left(
\frac{x}{\sqrt{2}\xi}\right)  ,\label{P2se}\\
P_{3}^{s} &  =\frac{\pi}{8\xi^{4}}K_{0}(\overline{\omega}r)\cos\left(
\frac{x}{\sqrt{2}\xi}\right)  \nonumber\\
&  \quad+\frac{\pi}{2}\int_{0}^{1}dye^{-xy}\frac{y^{3}}{1+2\xi^{2}y^{2}}%
I_{0}(\overline{\omega}_{3}r),\label{P3se}\\
P_{4}^{s} &  =\frac{\pi x}{2r^{3}\left(  1+u^{2}\right)  ^{3/2}}=\frac{\pi
x}{2\left(  x^{2}+r^{2}\right)  ^{3/2}}.\label{P4se}%
\end{align}
Collecting these results, we obtain the nonequilibrium pressure profile:%
\begin{align}
\Delta\widehat{p} &  =-\frac{\pi x}{2\left(  x^{2}+r^{2}\right)  ^{3/2}%
}\nonumber\\
&  \quad-\frac{\pi}{4}\int_{0}^{\sqrt{2\left(  1+\xi^{2}\right)  }}%
dy\frac{e^{-xy}y\left(  1+\sqrt{2}\xi y+\sqrt{1+2\xi^{2}y^{2}}\right)
}{1+2\xi^{2}y^{2}}I_{0}(\overline{\omega}_{1}r)\nonumber\\
&  \quad+\frac{\pi}{2}\int_{0}^{1}dy\frac{e^{-xy}\sqrt{2}\xi y^{2}\left(
1-\sqrt{2}\xi y\right)  }{1+2\xi^{2}y^{2}}I_{0}(\overline{\omega}%
_{3}r),\label{dpfin}%
\end{align}
This shows that $\Delta\widehat{p}$ is also singular at the origin of the
coordinates. It is significant to observe that the nonequilibrium pressure
$\Delta p$ is generally negative, that is, there is a tension that becomes
negative infinite at the origin. This means the nonequilibrium pressure is
compressional. It seems to be a remarkable result, probably deserving a deeper
consideration. We will report on a further study of this nonequilibrium
pressure separately.

By this, we have now shown that the formal solution of the NS equation of a
binary strong electrolyte in an external electric field can be expressed in
terms of elementary functions and well-behaved quadratures of regular Bessel
functions of second kind $I_{\nu}\left(  \overline{\omega}_{l}r\right)  $
$\left(  l=1,3\right)  $ for all values of $x$ and $r$. These results show how
the axial and transversal (radial) velocity and the nonequilibrium pressure
are distributed in the ($x,r$) space. In other words, they are velocity and
pressure distributions around the ions flowing in the medium subjected to an
external electric field of an arbitrary strength. They are hitherto unknown
results in hydrodynamics of electrolyte solutions. By using these profiles of
velocities and nonequilibrium pressure we will be able to deduce, in a
well-defined manner, numerous hydrodynamic consequences of ionic motions in
the medium in an electric field, subjected to irreversible thermodynamic principles.

\subsection{Transformation of Velocity to Spherical Coordinates}

In this work, the NS equation has been solved in cylindrical coordinates for
the reason that the external field has an axial symmetry, and hence the
velocity and pressure formulas are given in cylindrical coordinates. On the
other hand, the ions are regarded as either point charges or hard spheres with
spherical symmetry. To examine the motion of the medium around the charges
(e.g., hard spheres) it is convenient to have the velocity expressed in
coordinates adapted to the symmetry of ions, namely, spherical coordinates.
Therefore, it is necessary to transform the velocity vector in cylindrical
coordinates to that expressed in spherical coordinates. This aim can be
achieved with appropriate transformations for the velocity in cylindrical
coordinates $\left(  x,r,\theta\right)  $ to that in spherical coordinates
$\left(  R,\vartheta,\varphi\right)  $, which are related to each other by the
equations%
\begin{align}
\theta &  =\varphi,\nonumber\\
x  &  =R\cos\vartheta,\label{sc3}\\
r  &  =R\sin\vartheta,\nonumber
\end{align}
where $R$ is the radial coordinate, $\vartheta$ is the polar angle, and
$\varphi$ is the azimuthal angle of the spherical coordinate system. It should
be noted that the roles of $x$ and $z$ are switched from the conventional
usage. Accompanying these transformations, the unit vectors are related to
each other as follows:%
\begin{align}%
%TCIMACRO{\TeXButton{delta}{\mbox{\boldmath$\delta$}}}%
%BeginExpansion
\mbox{\boldmath$\delta$}%
%EndExpansion
_{x}  &  =\cos\vartheta%
%TCIMACRO{\TeXButton{delta}{\mbox{\boldmath$\delta$}}}%
%BeginExpansion
\mbox{\boldmath$\delta$}%
%EndExpansion
_{R}-\sin\vartheta%
%TCIMACRO{\TeXButton{delta}{\mbox{\boldmath$\delta$}}}%
%BeginExpansion
\mbox{\boldmath$\delta$}%
%EndExpansion
_{\vartheta},\nonumber\\%
%TCIMACRO{\TeXButton{delta}{\mbox{\boldmath$\delta$}}}%
%BeginExpansion
\mbox{\boldmath$\delta$}%
%EndExpansion
_{r}  &  =\sin\vartheta%
%TCIMACRO{\TeXButton{delta}{\mbox{\boldmath$\delta$}}}%
%BeginExpansion
\mbox{\boldmath$\delta$}%
%EndExpansion
_{R}+\cos\vartheta%
%TCIMACRO{\TeXButton{delta}{\mbox{\boldmath$\delta$}}}%
%BeginExpansion
\mbox{\boldmath$\delta$}%
%EndExpansion
_{\vartheta},\label{sc4}\\%
%TCIMACRO{\TeXButton{delta}{\mbox{\boldmath$\delta$}}}%
%BeginExpansion
\mbox{\boldmath$\delta$}%
%EndExpansion
_{\theta}  &  =%
%TCIMACRO{\TeXButton{delta}{\mbox{\boldmath$\delta$}}}%
%BeginExpansion
\mbox{\boldmath$\delta$}%
%EndExpansion
_{\varphi},\nonumber
\end{align}
where $%
%TCIMACRO{\TeXButton{delta}{\mbox{\boldmath$\delta$}}}%
%BeginExpansion
\mbox{\boldmath$\delta$}%
%EndExpansion
_{x}$, $%
%TCIMACRO{\TeXButton{delta}{\mbox{\boldmath$\delta$}}}%
%BeginExpansion
\mbox{\boldmath$\delta$}%
%EndExpansion
_{r}$, and $%
%TCIMACRO{\TeXButton{delta}{\mbox{\boldmath$\delta$}}}%
%BeginExpansion
\mbox{\boldmath$\delta$}%
%EndExpansion
_{\theta}$ are unit vectors in cylindrical coordinates whereas $%
%TCIMACRO{\TeXButton{delta}{\mbox{\boldmath$\delta$}}}%
%BeginExpansion
\mbox{\boldmath$\delta$}%
%EndExpansion
_{R}$, $%
%TCIMACRO{\TeXButton{delta}{\mbox{\boldmath$\delta$}}}%
%BeginExpansion
\mbox{\boldmath$\delta$}%
%EndExpansion
_{\vartheta}$, and $%
%TCIMACRO{\TeXButton{delta}{\mbox{\boldmath$\delta$}}}%
%BeginExpansion
\mbox{\boldmath$\delta$}%
%EndExpansion
_{\varphi}$ are unit vectors in spherical coordinates corresponding to
coordinates $R$, $\vartheta$, and $\varphi$. Since the velocity is
decomposable in the two coordinate systems as%
\begin{align}
\widehat{\mathbf{v}}  &  =\widehat{\mathbf{v}}_{x}%
%TCIMACRO{\TeXButton{delta}{\mbox{\boldmath$\delta$}}}%
%BeginExpansion
\mbox{\boldmath$\delta$}%
%EndExpansion
_{x}+\widehat{\mathbf{v}}_{\rho}%
%TCIMACRO{\TeXButton{delta}{\mbox{\boldmath$\delta$}}}%
%BeginExpansion
\mbox{\boldmath$\delta$}%
%EndExpansion
_{r}+\left(  0\right)
%TCIMACRO{\TeXButton{delta}{\mbox{\boldmath$\delta$}}}%
%BeginExpansion
\mbox{\boldmath$\delta$}%
%EndExpansion
_{\theta}\quad\text{in cylindrical coordinates,}\label{sc5a}\\
&  =\widehat{\mathbf{v}}_{R}%
%TCIMACRO{\TeXButton{delta}{\mbox{\boldmath$\delta$}}}%
%BeginExpansion
\mbox{\boldmath$\delta$}%
%EndExpansion
_{R}+\widehat{\mathbf{v}}_{\vartheta}%
%TCIMACRO{\TeXButton{delta}{\mbox{\boldmath$\delta$}}}%
%BeginExpansion
\mbox{\boldmath$\delta$}%
%EndExpansion
_{\vartheta}+\left(  0\right)
%TCIMACRO{\TeXButton{delta}{\mbox{\boldmath$\delta$}}}%
%BeginExpansion
\mbox{\boldmath$\delta$}%
%EndExpansion
_{\varphi}\quad\text{in spherical coordinates,} \label{sc5b}%
\end{align}
where $\widehat{\mathbf{v}}_{R}$ and $\widehat{\mathbf{v}}_{\vartheta}$ are
velocity components in spherical coordinates, using the relations in Eq.
(\ref{sc4}) we obtain the relations between velocity components in two
coordinate systems:%
\begin{align}
\widehat{\mathbf{v}}_{R}  &  =\cos\vartheta\widehat{\mathbf{v}}_{x}%
+\sin\vartheta\widehat{\mathbf{v}}_{\rho},\label{sc6a}\\
\widehat{\mathbf{v}}_{\vartheta}  &  =-\sin\vartheta\widehat{\mathbf{v}}%
_{x}+\cos\vartheta\widehat{\mathbf{v}}_{\rho}. \label{sc6b}%
\end{align}
Thus multiplying $\widehat{\mathbf{v}}_{R}$ with $\left(  -2\sqrt{2}\xi
/\pi\right)  $
\begin{equation}
\mathbf{v}_{R}=-\frac{2\sqrt{2}\xi}{\pi}\widehat{\mathbf{v}}_{R}, \label{sc7a}%
\end{equation}
we obtain the radial component of the velocity in spherical coordinates:%
\begin{align}
\mathbf{v}_{R}  &  =\frac{\left(  \cos\vartheta+\sin\vartheta\right)
\sin\vartheta}{R^{2}}+\frac{\sqrt{2}\xi}{R}\left[  \cos\vartheta\left(
1-\frac{\sin^{2}\vartheta}{2}\right)  +\frac{\sin^{2}\vartheta}{\sqrt{2}%
R}\right. \nonumber\\
&  \quad\left.  -\frac{\cos\vartheta\left(  2\cos^{2}\vartheta-\sin
^{2}\vartheta\right)  }{R^{2}}+\frac{3\sin\vartheta\left(  \cos^{4}%
\vartheta-\cos^{2}\vartheta\sin^{2}\vartheta-\sin^{4}\vartheta\right)  }%
{R^{5}}\right] \nonumber\\
&  \quad+2\sqrt{2}\xi\left[  \cos\vartheta\Theta\left(  R,\vartheta
;\xi\right)  -\sin\vartheta\Theta_{t}\left(  R,\vartheta,\xi\right)  \right]
, \label{sc7}%
\end{align}
where%
\begin{align}
\Theta &  =\frac{1}{4}\int_{0}^{\sqrt{2\left(  1+\xi^{2}\right)  }}%
dy\frac{e^{-yR\cos\vartheta}\left(  1-y^{2}+\sqrt{1+2\xi^{2}y^{2}}\right)
}{1+2\xi^{2}y^{2}}\times\nonumber\\
&  \qquad\qquad\qquad\qquad\quad\left[  1-\frac{\sqrt{2}\xi y}{\left(
1+\sqrt{1+2\xi^{2}y^{2}}\right)  }\right]  I_{0}(\overline{\omega}_{1}%
R\sin\vartheta)\nonumber\\
&  \qquad+\frac{1}{2}\int_{0}^{1}dy\frac{e^{-yR\cos\vartheta}\sqrt{2}\xi
y\left(  1-y^{2}\right)  \left(  \sqrt{2}\xi y-1\right)  }{1+2\xi^{2}y^{2}%
}I_{0}(\overline{\omega}_{3}R\sin\vartheta), \label{sc9}%
\end{align}%
\begin{align}
\Theta_{t}  &  =\frac{1}{4}\int_{0}^{\sqrt{2\left(  1+\xi^{2}\right)  }%
}dy\frac{e^{-yR\cos\vartheta}y\left(  1-y^{2}+\sqrt{1+2\xi^{2}y^{2}}\right)
}{\left(  1+2\xi^{2}y^{2}\right)  }\times\nonumber\\
&  \qquad\qquad\qquad\qquad\left(  1+\frac{\xi y}{\sqrt{2}\left(
1+\sqrt{1+2\xi^{2}y^{2}}\right)  }\right)  I_{1}(\overline{\omega}_{1}%
R\sin\vartheta)\nonumber\\
&  \qquad+\xi^{2}\int_{0}^{1}dy\frac{e^{-yR\cos\vartheta}y^{2}\left(
y-\sqrt{2}\xi\right)  \left(  1-y^{2}\right)  }{1+2\xi^{2}y^{2}}%
I_{1}(\overline{\omega}_{3}R\sin\vartheta). \label{sc10}%
\end{align}
These formulas may be used to calculate the force on the center ion of ion
atmosphere in the external field. It is closely related to the electrophoretic
effect we will need for study of conductance.

In Fig. 5 and Fig. 6, the radial velocity profile $\mathbf{v}_{R}$ is
presented for two cases of $\xi$ to show the behavior of the countercurrent in
spherical coordinates. In this contour map of $\mathbf{v}_{R}$, its magnitude
decreases as the color changes from red to blue according to the scale shown
on the right in the figures. It is, in fact, divergent at the origin, meaning
an infinite countercurrent, and gradually decreases along the axial direction
$\left(  \vartheta=0\right)  $, but as the angle $\vartheta$ increases beyond
$\vartheta\approx40$ in the case of, for example, $\xi=0.1$ it becomes
negative when $R\lesssim1$ as is evident from the right-hand corner of Fig. 5.
This behavior of the countercurrent does not basically change as the field
strength $\xi$ increases as is evident from Fig. 6 for $\xi=3.0$. The
negativity of $\mathbf{v}_{R}$ arises from the deterministic part $\left(
\mathbf{v}\right)  _{\text{acme}}$ defined by Eq. (\ref{vacm}), which can be
negative, but the Brownian motion part of the contribution $\left[
\cos\vartheta\Theta\left(  R,\vartheta;\xi\right)  -\sin\vartheta\Theta
_{t}\left(  R,\vartheta,\xi\right)  \right]  $ is everywhere positive. This
feature is shown in Fig. 7 and Fig. 8, in which the contour maps show
structures, but do not show a negative domain. The color code bar on the right
of the figures is for the magnitude of the function plotted. In Fig. 9, we
show the variation of $\Theta\left(  R,0;\xi\right)  $ with respect to $R$ and
$\xi$, showing its decreasing tendency with increasing $R$ and also with $\xi$
at a given value of $R$. These figures qualitatively show in which region of
$\left(  R,\vartheta\right)  $ the countercurrent is least in magnitude, thus
offering a least resistance to the motion of charges pulled by the external
electric field. This feature therefore offers a useful insight into studying
mobility of ions in the electrolyte solutions in the external electric field
that, hopefully,\ can be made use of in the light of irreversible
thermodynamic principles..

\section{\textbf{Relation to the Result by Onsager and Wilson}}

Having obtained an axial velocity formula as a function of $x$ and $r$, it is
appropriate to see how the axial velocity obtained by Onsager and
Wilson\cite{wilson} for a special position of $x=r=0$, namely, the coordinate
origin at which the center ion of the ion atmosphere is placed, may be
recovered. In particular, it is important to see in what manner Wilson's
formula for the electrophoretic effect, $f\left(  \xi\right)  $, defined by
the relation%
\begin{equation}
v_{x}\left(  0,0,0\right)  =-\frac{zeX\kappa}{6\sqrt{2}\pi\eta_{0}}f\left(
\xi\right)  , \label{fW}%
\end{equation}
should be understood, given the exact velocity profile formula (\ref{rvax}).
Here $v_{x}\left(  0,0,0\right)  $ is the axial velocity of ion at the origin
obtained by evaluating the integrals in the\ axial velocity formula
(\ref{38}), or its reduced form in Eq. (\ref{rv}) with Eqs. (\ref{Kc}) and
(\ref{Ks}), on taking $x=0$ and $r=0$ in the integrals.

At $x=r=0$ in Eq. (\ref{38}), the sine transforms all vanish and the cosine
transforms contribute well-behaved integrals except for the integral
$K_{5}^{c}$ [Eq. (\ref{Kc5})], which gives rise to a divergent integral, Eq.
(\ref{divi}), as pointed out in Subsec. III.4. In Wilson's
dissertation\cite{wilson}, provided that the integral $K_{5}^{c}$ is ignored,
the electrophoretic effect $f\left(  \xi\right)  $ is found given by%
\begin{align}
f\left(  \xi\right)   &  =1+\frac{3}{4\sqrt{2}\xi^{3}}\left\{  2\xi^{2}%
\sinh^{-1}\xi+\sqrt{2}\xi-\xi\sqrt{1+\xi^{2}}\right.  \nonumber\\
&  \quad\left.  -\left(  1+2\xi^{2}\right)  \tan^{-1}\left(  \sqrt{2}%
\xi\right)  +\left(  1+2\xi^{2}\right)  \tan^{-1}\left(  \frac{\xi}%
{\sqrt{1+\xi^{2}}}\right)  \right\}  .\label{efW}%
\end{align}
This formula for the electrophoretic effect is in the foundation of the theory
of Wien effect for binary strong electrolytes\cite{harned}. In view of the
divergent integral mentioned, we would like to see in which manner this
formula should be obtained from the general formula for the axial velocity,
Eq. (\ref{rvax}),\ that we have derived from the solution of the NS equation.

The difficulty of getting this formula from Eq. (\ref{rvax}) resides in the
term related to $\left(  \widehat{\mathbf{v}}_{x}\right)  _{\text{acme}}$ [Eq.
(\ref{vacm})]%
\begin{equation}
\Phi\left(  x,r;\xi\right)  =\frac{x}{\left(  x^{2}+r^{2}\right)  ^{3/2}%
}+\sqrt{2}\xi\left[  \frac{1}{\left(  x^{2}+r^{2}\right)  ^{1/2}}-\frac{r^{2}%
}{2\left(  x^{2}+r^{2}\right)  ^{\frac{3}{2}}}-\frac{2x^{2}-r^{2}}{\left(
x^{2}+r^{2}\right)  ^{5/2}}\right]  , \label{phi}%
\end{equation}
which we have termed the \textquotedblleft deterministic\textquotedblright%
\ part of the axial velocity ($-\left(  \widehat{\mathbf{v}}_{x}\right)
_{\text{acme}}$)---a terminology given to it because it originates from the
non-dissipative part of the body force in contrast to the other terms in the
axial velocity, that is, the quadratures in Eq. (\ref{rvax}). We have
multiplied $\sqrt{2}\xi/\pi$ to this factor because the field strength $X$
must be multiplied to obtain the axial velocity. This function $\Phi\left(
x,r;\xi\right)  $ clearly diverges at the origin of the coordinate system. In
other words, the axial velocity profiles never passes through the origin at
any value of the field strength $\xi$ because the coordinate origin is a
singular point of $\widehat{v}_{x}\left(  x,r,0\right)  $; in other words, the
countercurrent is infinite at the origin. Furthermore, by examining the upper
bounds of the quadratures, the quadratures in $\widehat{v}_{x}$ exponentially
increase with respect to $\xi$ if $x$ is such that $xy\leq\overline{\omega
}_{1}r$ in the interval $\left[  1,\sqrt{2\left(  1+\xi^{2}\right)  }\right]
$\ in the large $\xi$ (field strength) limit. This would make the
countercurrent very large in magnitude in the interval and thus renders the
ionic flow unrealizable. However, there is a set of trajectories of ($x,r$)
for which $xy\geq\overline{\omega}_{1}r$ and thus the quadratures are
exponentially decreasing as $\xi$ increases; these figures are not shown in
this paper for lack of space, but the velocity profiles are instead shown in
spherical coordinates in the previous section. Therefore the optimum value of
$\left(  x,r\right)  $ must be that of $\min\left[  \widehat{v}_{x}\left(
x,r;\xi\right)  \right]  $ for a given field strength, namely, the trajectory
of $(x,r)$ corresponding to the valley of the axial velocity surface. This
trajectory $(x,r)_{x=x_{c},r=r_{c}}\in\min\left[  \widehat{v}_{x}\left(
x,r;\xi\right)  \right]  $ can be determined as a function of $\xi$ by a
numerical means only. The values of $x_{c}$ and $r_{c}$ are $O(1)$ in reduced
units of distance or $\kappa^{-1}$ in actual units. We have already shown in
the previous section that this aspect is better displayed if spherical
coordinates are employed.

In any case, Wilson's result for $f\left(  \xi\right)  $ can be recovered,
\textit{only (1) if the factor }$\Phi\left(  x,r;\xi\right)  $\textit{ is
ignored or the arguments }$x$\textit{ and }$r$\textit{ are chosen so that
}$\Phi\left(  x,r;\xi\right)  =0$\textit{ in the definition of electrophoretic
effect and (2) if not only }$x$\textit{ and }$r$\textit{ are set equal to zero
in the integrals in Eq. (\ref{rvax}) for }$\widehat{v}_{x}$\textit{, but also
the terms therein that arise from the sine transforms are omitted, that is,
the second term in the square brackets in Eq. (\ref{theta}) given below.} We
explain it more explicitly: First, if we set $x=r=0$ in the integrals of Eq.
(\ref{rvax}) we obtain
\begin{align}
\Xi &  \equiv-\frac{\pi}{4}\int_{0}^{\sqrt{2\left(  1+\xi^{2}\right)  }%
}dy\frac{\left(  1-y^{2}+\sqrt{1+2\xi^{2}y^{2}}\right)  }{1+2\xi^{2}y^{2}%
}\left[  1-\frac{\sqrt{2}\xi y}{\left(  1+\sqrt{1+2\xi^{2}y^{2}}\right)
}\right] \nonumber\\
&  \qquad-\frac{\pi\xi}{\sqrt{2}}\int_{0}^{1}dy\frac{y\left(  1-y^{2}\right)
}{1+2\xi^{2}y^{2}}\left[  \sqrt{2}\xi y-1\right]  . \label{theta}%
\end{align}
For this we note that%
\[
I_{0}\left(  z\right)  |_{z=0}=1.
\]
Furthermore, if the second term in the square brackets in the first and second
integrals in $\Xi$ is neglected, that is, the terms $\sqrt{2}\xi y/\left(
1+\sqrt{1+2\xi y^{2}}\right)  $ and $1$, the resulting integrals are precisely
the integrals Wilson\cite{wilson} evaluated to obtain the result for $f\left(
\xi\right)  $ given in Eq. (\ref{efW}):%
\begin{equation}
\Xi_{W}=-\frac{\pi}{2}\left[  \frac{1}{2}\int_{0}^{\sqrt{2\left(  1+\xi
^{2}\right)  }}dy\frac{\left(  1-y^{2}+\sqrt{1+2\xi^{2}y^{2}}\right)  }%
{1+2\xi^{2}y^{2}}+2\xi^{2}\int_{0}^{1}dy\frac{y^{2}\left(  1-y^{2}\right)
}{1+2\xi^{2}y^{2}}\right]  . \label{thetaW}%
\end{equation}
The integrals here are elementary. It is easy to show that $\Xi_{W}$ indeed
gives rise to $f\left(  \xi\right)  $ in Eq. (\ref{efW}).

By this, we see that the connection of $f\left(  \xi\right)  $ to the axial
velocity formula in Eq. (\ref{rvax}) can be made only when we arbitrarily set
$\Phi=0$ and also $x=0$ and $r=0$ in the integrals and when some terms therein
originating from the sine transform terms in the axial velocity are dropped arbitrarily.

In any case, in view of the fact that the Onsager--Wilson theory for Wien
effect predicts ionic conductances too small in comparison\cite{eckstrom} with
experiment it would be useful to examine whether the complete velocity profile
formula would improve the theoretical prediction. Application to the question
of electrophoretic factor of the axial velocity formula will be made in the
sequel\cite{sequel} where conductivity will be studied in comparison with experiment.

\section{\textbf{Discussion and Concluding Remarks}}

Since ions interact with each other through long range Coulombic interactions,
and ion atmospheres with ions and the external field, the whole body of the
solution collectively and cooperatively moves subjected to the external field.
Consequently, the macroscopic behavior of electrolyte solutions in the
presence of an external electric field is not simple.

Their electrical conductivity in particular has attracted a great deal of
attention in physical chemistry for reasons related to the basic questions in
chemistry and thermodynamics including the behaviors of electrolytic solutions
from the early days of physical chemistry. On the basis of Debye's theory of
electrolytic solutions\cite{debye}, Lars Onsager\cite{onsager,onsager3}, in
particular, elucidated the physical mechanisms---the electrophoretic effect
and the relaxation time effect---underlying the conduction phenomena, which
ultimately require solutions of hydrodynamic equations together with
appropriate constitutive equations for irreversible nonequilibrium fluxes in
terms of thermodynamic forces. The Onsager--Wilson (OW) theory of Wien effect
\cite{wilson}, and Onsager--Kim (OK) theory\cite{kim} in the case of
asymmetric strong electrolytes, depicts the manner in which nonlinear field
effects on ionic conductivity and mobility can be studied. However, critical
studies of these theories have not been made in the literature. Even numerical
studies of, for example, the OK theory, although difficult, have not been made
as yet. The Onsager--Liu theory \cite{liu} and the Onsager--Chen theory
\cite{chen} were the most recent follow-ups in the aforementioned line, but
their works were concerned with higher-order density corrections, not the Wien
effect itself.

The ionic association theory approach by Patterson and his
collaborators\cite{patterson} using the Bjerrum's theory\cite{bjerrum} of
ionic association and Onsager's theory of weak electrolytic
conduction\cite{onsager3} was based on the electrophoretic effect obtained by
Wilson despite the question of divergence mentioned earlier. In view of
Wilson's electrophoretic effect factor $f(\xi)$ needing reassessment and
probably a revision because of the divergence difficulty mentioned, and the
notion of ionic association that probably can be better addressed by means of
the modern statistical mechanics of ionic liquids\cite{friedman,blum}, the
approach of Patterson et al. would require a fresh re-examination. Therefore,
the theory of Wien effect and more generally, nonlinear field effects on ionic
conductivity of electrolyte solutions is not a closed subject as yet despite
its long history in the opinion of the present authors.

At this point, it is useful to reconsider, by using the radial velocity
component $\widehat{\mathbf{v}}_{R}$ given in Eq. (\ref{sc7}), the connection
of the present work to the result for the electrophoretic effect examined with
$\widehat{\mathbf{v}}_{x}$ in cylindrical coordinate in Sec. V. Here we would
like to consider the radial velocity $\widehat{\mathbf{v}}_{R}$ in the
direction parallel to the external field, that is, $\vartheta=0$, which then
reads
\begin{equation}
\mathbf{v}_{R}\left(  R,0,\xi\right)  =\frac{\sqrt{2}}{R}\left(  1-\frac
{2}{R^{2}}\right)  +2\sqrt{2}\Theta\left(  R,0;\xi\right)  . \label{sc8}%
\end{equation}
Before proceeding further, it is useful to note that if we define a
generalized electrophoretic factor $\mathfrak{f}\left(  R,\vartheta
,\xi\right)  $ on the basis of the velocity in spherical coordinates
$\mathbf{v}\left(  R,\vartheta,\varphi\right)  \equiv\mathbf{v}_{x}%
\cos\vartheta+\mathbf{v}_{\rho}\sin\vartheta$ in analogy to Eq. (\ref{fW})%
\begin{equation}
\mathbf{v}\left(  R,\vartheta,\varphi\right)  =-\frac{zeX\kappa}{6\sqrt{2}%
\pi\eta_{0}}\mathfrak{f}\left(  R,\vartheta,\xi\right)  , \label{sc8f}%
\end{equation}
then $\mathbf{v}_{R}\left(  R,0,\xi\right)  $ is related to $\mathfrak{f}%
\left(  R,0,\xi\right)  \mathfrak{\ }$by the relation
\begin{equation}
\mathbf{v}_{R}\left(  R,0,\xi\right)  /\xi=\frac{2\sqrt{2}}{3}\mathfrak{f}%
\left(  R,0,\xi\right)  . \label{scf0}%
\end{equation}
If $R=\sqrt{2},\vartheta=0$ then%
\begin{equation}
\mathbf{v}_{R}\left(  \sqrt{2},0,\xi\right)  /\xi=\frac{2\sqrt{2}}%
{3}\mathfrak{f}\left(  \sqrt{2},0;\xi\right)  =2\sqrt{2}\Theta\left(  \sqrt
{2},0;\xi\right)  . \label{sc8a}%
\end{equation}
Note that in the actual distance scale $R=\sqrt{2}$ corresponds to the radial
position at $2\kappa^{-1}$, i.e., twice the Debye length, and at this point
the factor $\Phi$ [Eq. (\ref{phi})] related to $\left(  \widehat{\mathbf{v}%
}_{x}\right)  _{\text{acme}}$, the deterministic part of the velocity,
vanishes: $\left(  \widehat{\mathbf{v}}_{x}\right)  _{\text{acme}}%
|_{R=\sqrt{2},\vartheta=0}=0$. And the choice of this particular value of
$R=\sqrt{2}$ replaces the step by which the divergent integral was arbitrarily
discarded in Wilson's procedure. Then the resulting $\mathfrak{f}\left(
\sqrt{2},0,\xi\right)  $ is rather a similar to $f\left(  \xi\right)  $, which
can be shown given by $\Xi_{W}$ (Eq. (\ref{thetaW})). We emphasize that here
the choice of $R=\sqrt{2}$ is guided by $\Phi\left(  R,\vartheta\right)
|_{R=\sqrt{2},\vartheta=0}=0$. We will explore its possible irreversible
thermodynamics basis in the sequel.

On the other hand, If $\vartheta=0$ and $R=1/\sqrt{2}$, i.e., at$\sqrt{2}$
times the Debye length $\kappa^{-1}$ of the radial position,%
\begin{equation}
\mathbf{v}_{R}\left(  1/\sqrt{2},0,\xi\right)  /\xi=\frac{2\sqrt{2}}%
{3}\mathfrak{f}\left(  1/\sqrt{2},0,\xi\right)  =\sqrt{2}\left(
2\Theta\left(  1/\sqrt{2},0;\xi\right)  -1\right)  . \label{sc8b}%
\end{equation}
These two cases are plotted In Fig. 10 and Fig. 11, respectively, with respect
to $\xi$. We notice that $\mathbf{v}_{R}\left(  1/\sqrt{2},0,\xi\right)
/\xi>>\mathbf{v}_{R}\left(  \sqrt{2},0,\xi\right)  /\xi$, or, put in another
way, the countercurrent is much larger in magnitude in the case of
$R=1/\sqrt{2}$ than $R=\sqrt{2}$.

These two cases of $\mathfrak{f}\left(  R,0,\xi\right)  $ correspond to
$f\left(  \xi\right)  $ in Wilson's work, but not equal to $f\left(
\xi\right)  $. Evidently, they are well-behaved and finite functions of $\xi$,
whose behavior is reminiscent of that of $f(\xi)$. Interestingly,
$\mathfrak{f}\left(  \sqrt{2},0,\xi\right)  $ is closer in magnitude to
$f\left(  \xi\right)  $ than $\mathfrak{f}\left(  1/\sqrt{2},0,\xi\right)  $.
This aspect and the related will be given a more detailed consideration in the
sequel\cite{sequel} where application of the results of the present work to
the theory of conductance will be examined in detail with the help of the
irreversible thermodynamic principles associated with the flow.

The present work was born out of our desire to fully comprehend the meanings
of the divergence-causing integral(s) in the formal Fourier transform solution
of the NS equation of binary electrolytes in the presence of an external
electric field and to find thereby a way to circumvent the divergence in
question. In this article, to achieve this goal we have fully evaluated the
aforementioned formal solution of the NS equation without assuming a special
case of spatial positions, in terms of simple elementary functions plus
quadratures of well-behaved functions. What we have obtained are the exact
velocity and pressure profiles in space, which we may apply to study
irreversible phenomena in the binary electrolyte solutions in the electric
field, including electrical conduction phenomena. Being a full exact solution
without an approximation, it promises to provide a more complete picture of
conduction phenomena. This part of the study will be made in the sequels to
this article.

The solution [Eqs. (\ref{rvax}), (\ref{rvrfin}), (\ref{dpfin}), and
(\ref{sc7})] of the NS equation obtained here is a rare example of rather
simple, but explicit solutions for the kind of complex, but practical systems,
that are amenable to relatively straightforward mathematical analysis and
computation by a desktop computer, because the formulas involved are either
elementary functions or comparatively simple quadratures free from a singular
behavior, such as of poles and other singularities. They therefore appear to
be potentially very useful for gaining insights into and describing how ions
in electrolyte solutions move in the presence of an external electric field.
These insights also should help us develop theories of electrical conductivity
and related transport phenomena in systems\cite{ting}$^{-}$\cite{nguyen} of
current interest in science and engineering, such as plasmas, semiconductors,
etc. in electromagnetic fields.

\bigskip

\textbf{Acknowledgment}

\textit{The present work has been supported in part by the Discovery grants
from the Natural Sciences and Engineering Research Council of Canada.}

\appendix{}

\section{Examples for Contour Integration of Integrals}

In this Appendix, we use Integrals $K_{1}^{c}$ and $K_{1}^{s}$ [Eqs.
(\ref{Kc}) and (\ref{Ks})] as typical examples illustrating the methods of
contour integration used in this work. Other integrals can be evaluated
similarly by using appropriate contours given in Fig. 3 and Fig. 4.

\subsection{Integrals $K_{1}^{c}$ and $K_{1}^{s}$}

Fig. 2 is used for evaluating this integral. Consider the contour integral
along contour $\mathcal{C}_{1}$ in complex plane $z$ depicted in Fig. 2
\begin{equation}
\mathcal{C}_{1}K_{1}^{c}=\int_{\mathcal{C}_{1}}dze^{ixz}\frac{\omega_{1}^{2}%
}{1-2\xi^{2}z^{2}}K_{0}(\omega_{1}r), \label{A2}%
\end{equation}
where
\begin{equation}
\omega_{1}=\left[  1+z^{2}+\sqrt{1-2\xi^{2}z^{2}}\right]  ^{1/2}. \label{A3}%
\end{equation}
Since there is no singularity enclosed by the contour, this contour integral
$\mathcal{C}_{1}K_{1}^{c}$ is clearly equal to zero. Integral $\mathcal{C}%
_{1}K_{1}^{c}$ can be decomposed into integrals along the paths $C_{-}$,
$C_{+}$, $C$, $C_{\infty}$, and along the real axis $t$. We thus may write it
as%
\begin{align}
\mathcal{C}_{1}K_{1}^{c}  &  =\int_{-\infty}^{\infty}dt\frac{\left(
1+t^{2}+\sqrt{1-2\xi^{2}t^{2}}\right)  }{1-2\xi^{2}t^{2}}e^{ixt}K_{0}%
(\omega_{1}\left(  t\right)  r)\nonumber\\
&  +\int_{C_{-}}dz\frac{\left(  1+z^{2}+\sqrt{1-2\xi^{2}z^{2}}\right)
}{1-2\xi^{2}z^{2}}e^{ixz}K_{0}(\omega_{1}\left(  z\right)  r)\nonumber\\
&  +\int_{C_{+}}dz\frac{\left(  1+z^{2}+\sqrt{1-2\xi^{2}z^{2}}\right)
}{1-2\xi^{2}z^{2}}e^{ixz}K_{0}(\omega_{1}\left(  z\right)  r)\nonumber\\
&  +\int_{C}dz\frac{\left(  1+z^{2}+\sqrt{1-2\xi^{2}z^{2}}\right)  }%
{1-2\xi^{2}z^{2}}e^{ixz}K_{0}(\omega_{1}\left(  z\right)  r)\nonumber\\
&  +\int_{C_{\infty}}dz\frac{\left(  1+z^{2}+\sqrt{1-2\xi^{2}z^{2}}\right)
}{1-2\xi^{2}z^{2}}e^{ixz}K_{0}(\omega_{1}\left(  z\right)  r)\nonumber\\
&  =0. \label{A4}%
\end{align}
The first integral on the right can be shown to be equal to $2K_{1}^{c}$. By
the theorem of residues\cite{whittaker} the integrals $C_{-}K_{1}^{c}$ and
$C_{+}K_{1}^{c}$ gives $\pi i$ times residue of $\left(  C_{\pm}K_{1}%
^{c}\right)  $. Thus we obtain%
\begin{equation}
C_{-}K_{1}^{c}+C_{+}K_{1}^{c}=\frac{\sqrt{2}\pi\left(  1+2\xi^{2}\right)
}{4\xi^{3}}\sin\left(  \frac{x}{\sqrt{2}\xi}\right)  K_{0}\left(
\overline{\omega}r\right)  , \label{A5}%
\end{equation}
where%
\begin{equation}
\overline{\omega}=\frac{\sqrt{1+2\xi^{2}}}{\sqrt{2}\xi}. \label{A6}%
\end{equation}
To transform the contour integral $CK_{1}^{c}$ around the branch cut along the
imaginary axis we observe that if the phase of the argument of $K_{0}\left(
\omega_{1}r\right)  $ on the right hand lip of the cut is chosen equal to zero
the phase of the argument on the left hand lip is $\pi i$, so that the
argument has the form $e^{\pi i}\omega_{1}r$ for the Bessel function on the
left side of contour $C$. We now observe that the Bessel function can be
continued from the left side of the cut to the right side by the following
continuation formula\cite{watson}%
\begin{equation}
K_{0}\left(  e^{i\pi}z\right)  =K_{0}\left(  z\right)  -\pi iI_{0}\left(
z\right)  , \label{A7}%
\end{equation}
where $I_{0}\left(  z\right)  $ is the regular solution for the second kind of
the Bessel function of order $0$, $K_{0}(z)$ being irregular in contrast to
$I_{0}\left(  z\right)  $ being regular. The irregular Bessel function
$K_{\nu}(z)$ $\left(  \nu\geq0\right)  $ diverges logarithmically as
$z\rightarrow0$. In series representation the Bessel function $I_{0}\left(
z\right)  $ is given by the formula%
\begin{equation}
I_{0}\left(  z\right)  =\sum_{m=0}^{\infty}\frac{\left(  \frac{1}{2}z\right)
^{2m}}{\left(  m!\right)  ^{2}}. \label{A8}%
\end{equation}
This function is finite at $z=0$, but it behaves asymptotically as%
\begin{equation}
I_{0}\left(  z\right)  \sim\left(  2\pi z\right)  ^{-1/2}e^{z}\left[
1+O(z^{-1})\right]  \quad\left(  \left\vert \arg z\right\vert <\right)  .
\label{A8a}%
\end{equation}
Using formula (\ref{A7}) and changing variable from $iy$ to $y$, we obtain%
\begin{equation}
CK_{1}^{c}=\pi\int_{0}^{\sqrt{2\left(  1+\xi^{2}\right)  }}dy\frac
{e^{-xy}\left(  1-y^{2}+\sqrt{1+2\xi^{2}y^{2}}\right)  }{1+2\xi^{2}y^{2}}%
I_{0}\left(  \overline{\omega}_{1}r\right)  , \label{A9}%
\end{equation}
where%
\begin{equation}
\overline{\omega}_{1}=\left[  1-y^{2}+\sqrt{1+2\xi^{2}y^{2}}\right]  ^{1/2}.
\label{A9a}%
\end{equation}
The contour integral along the infinite semicircle $C_{\infty}$ vanishes
identically because $K_{0}\left(  \omega_{1}r\right)  $
vanishes\cite{watson,abramowitz} as $\left\vert z\right\vert \rightarrow
\infty$.\ Collecting the results obtained above into Eq. (\ref{A4}), we
evaluate the integral $K_{1}^{c}$ in the form:%
\begin{align}
K_{1}^{c}  &  =-\frac{\sqrt{2}\pi\left(  1+2\xi^{2}\right)  }{8\xi^{3}}%
\sin\left(  \frac{x}{\sqrt{2}\xi}\right)  K_{0}\left(  \overline{\omega
}r\right) \nonumber\\
&  \quad\;-\frac{\pi}{2}\int_{0}^{\sqrt{2\left(  1+\xi^{2}\right)  }}%
dy\frac{e^{-xy}\left(  1-y^{2}+\sqrt{1+2\xi^{2}y^{2}}\right)  }{1+2\xi
^{2}y^{2}}I_{0}\left(  \overline{\omega}_{1}r\right)  . \label{A10}%
\end{align}

The procedure of evaluating integrals $K_{1}^{s}$ is entirely parallel to the
one presented above for $K_{1}^{c}$ with the contour in Fig. 2. The result for
the reduced integral is $K_{1}^{s}$%
\begin{align}
K_{1}^{s}  &  =\frac{\pi}{8\xi^{4}}\left(  1+2\xi^{2}\right)  \cos\left(
x/\sqrt{2}\xi\right)  K_{0}\left(  \omega_{1}\left(  -1/\sqrt{2}\xi\right)
r)\right) \nonumber\\
&  \quad\;-\frac{\pi}{2}\int_{0}^{\sqrt{2\left(  1+\xi^{2}\right)  }}%
dy\frac{e^{-xy}y\left(  1-y^{2}+\sqrt{1+2\xi^{2}y^{2}}\right)  }{\left(
1+2\xi^{2}y^{2}\right)  \left(  1+\sqrt{1+2\xi^{2}y^{2}}\right)  }I_{0}\left(
\overline{\omega}_{1}r\right)  . \label{A11}%
\end{align}

The procedures for integrals $K_{2}^{c}$, $K_{3}^{c}$, $K_{2}^{s}$, $K_{3}%
^{s}$ in integrals $K_{c}$ and $K_{s}$ are entirely parallel to those
presented here for $K_{1}^{c}$ and $K_{1}^{s}$ if the contours in Fig. 3--4
are made use of.

\subsection{Evaluation of Pair Distribution functions and Potentials}

The formal formulas for the pair distribution functions and potentials
presented in Sec. II can be evaluated by using the same methods as for the
velocity and pressure. In this part of Appendix A, we present the results for
them for completeness, although we have not needed them for the purpose of the
present paper.

\subsubsection{Pair distribution Functions}

With the definition of the symbol%
\begin{equation}
\Delta\widehat{f}_{kl}=\left(  \frac{n\kappa^{3}}{2\sqrt{2}\pi^{2}z}\right)
^{-1}\left(  f_{kl}-n^{2}\right)  \quad\left(  kl=ii,jj,ij\right)  \label{A12}%
\end{equation}
we obtain
\begin{align}
\Delta\widehat{f}_{ii}  &  =-\frac{\pi}{2}\int_{0}^{\sqrt{2\left(  1+\xi
^{2}\right)  }}dy\frac{e^{-xy}\left(  1-y^{2}+\sqrt{1+2\xi^{2}y^{2}}\right)
}{1+2\xi^{2}y^{2}}I_{0}\left(  \overline{\omega}_{1}r\right) \nonumber\\
&  -\frac{\pi}{2}\int_{0}^{1}dye^{-yx}\frac{4y^{2}\xi^{2}}{\left(  1+2\xi
^{2}y^{2}\right)  }I_{0}\left(  \overline{\omega}_{3}r\right)  , \label{A13}%
\end{align}%
\begin{align}
\Delta\widehat{f}_{jj}  &  =\frac{\sqrt{2}\pi}{\xi}\sin\left(  \frac{x}%
{\sqrt{2}\xi}\right)  K_{0}\left(  \overline{\omega}r\right) \nonumber\\
&  +\frac{\pi}{2}\int_{0}^{\sqrt{2\left(  1+\xi^{2}\right)  }}dy\frac
{e^{-xy}\left(  1-y^{2}+\sqrt{1+2\xi^{2}y^{2}}\right)  }{1+2\xi^{2}y^{2}}%
I_{0}\left(  \overline{\omega}_{1}r\right) \nonumber\\
&  -\frac{\pi}{2}\int_{0}^{1}dye^{-yx}\frac{4y^{2}\xi^{2}}{\left(  1+2\xi
^{2}y^{2}\right)  }I_{0}\left(  \overline{\omega}_{3}r\right)  , \label{A14}%
\end{align}
and%
\begin{align}
\Delta\widehat{f}_{ij}  &  =-\frac{\pi}{2}\int_{0}^{\sqrt{2\left(  1+\xi
^{2}\right)  }}dy\frac{e^{-xy}\left[  \left(  1+\sqrt{1+2\xi^{2}y^{2}}\right)
\left(  1\mp\frac{\xi}{\sqrt{2}}y\right)  -y^{2}\right]  }{1+2\xi^{2}y^{2}%
}I_{0}\left(  \overline{\omega}_{1}r\right) \nonumber\\
&  \qquad\pm\frac{\pi}{2}\frac{\xi}{\sqrt{2}}\int_{0}^{1}dyy\frac{2e^{-yx}%
}{\left(  1+2\xi^{2}y^{2}\right)  }I_{0}\left(  \overline{\omega}_{3}r\right)
. \label{A15}%
\end{align}

\subsubsection{Potentials}

With the definition of symbols%
\begin{equation}
\widehat{\psi}_{j}=\psi_{j}\left(  \pm\mathbf{r}\right)  \left(
\frac{ze\kappa}{\sqrt{2}\pi D}\right)  ^{-1}=-\psi_{i}\left(  \mp
\mathbf{r}\right)  \left(  \frac{ze\kappa}{\sqrt{2}\pi D}\right)  ^{-1}
\label{A16}%
\end{equation}
we obtain%
\begin{align}
\widehat{\psi}_{j}  &  =-\frac{\pi}{2}\int_{0}^{\sqrt{2\left(  1+\xi
^{2}\right)  }}dy\frac{ye^{-yx}I_{0}\left(  \overline{\omega}_{1}r\right)
}{\left(  1+2\xi^{2}y^{2}\right)  }\left(  \sqrt{1+2\xi^{2}y^{2}}+1\mp\sqrt
{2}\xi\right) \nonumber\\
&  -\pi\int_{0}^{1}dyy\frac{e^{-yx}}{\left(  1+2\xi^{2}y^{2}\right)  }%
I_{0}\left(  \overline{\omega}_{3}r\right)  \left(  1\mp\sqrt{2}\xi\right)  .
\label{A17}%
\end{align}

\newpage

Figure Captions

\bigskip

Fig. 1\quad The cylindrical coordinate system employed.

Fig. 2\quad Contour $\mathcal{C}_{1}$ for integrals $K_{1}^{c}$ and $K_{1}%
^{s}$. This contour also applies to integrals $J_{1}^{c}$ and $J_{1}^{s}$ and
$P_{1}^{c}$ and $P_{1}^{s}$. The bold line denotes the branch cut.

Fig. 3\quad Contour $\mathcal{C}_{2}$ for integrals $K_{2}^{c}$ and $K_{2}%
^{s}$. This contour also applies to integrals $J_{2}^{c}$ and $J_{2}^{s}$ and
$P_{2}^{c}$ and $P_{2}^{s}$. The bold line denotes the branch cut.

Fig. 4\quad Contour $\mathcal{C}_{3}$ for integrals $K_{3}^{c}$ and $K_{3}%
^{s}$. This contour also applies to integrals $J_{3}^{c}$ and $J_{3}^{s}$ and
$P_{3}^{c}$ and $P_{3}^{s}$. The bold line denotes the branch cut, which is on
the negative real axis.

Fig. 5\quad Contour maps of the radial velocity profile for different field
strength in the case of $\xi=0.1$. The radial velocity becomes singular at the
coordinate origin. The magnitude of velocity diminishes from red to blue.

Fig. 6\quad Contour maps of the radial velocity profile for different field
strength in the case of $\xi=3.0$. The feature of the figure is similar to
Fig. 5.

Fig. 7\quad The scaled radial velocity profile $\Theta\left(  R,\vartheta
,\xi\right)  $ in the case of $\xi=0.1$.

Fig. 8\quad The scaled radial velocity profile $\Theta\left(  R,\vartheta
,\xi\right)  $ in the case of $\xi=3.0$.

Fig. 9\quad The cross section of $\Theta\left(  R,\vartheta,\xi\right)  $ at
$\vartheta=0$ plotted against $R$ for different values of $\xi$. This figure
gives an idea of how the $\Theta\left(  R,\vartheta,\xi\right)  $ profiles
vary with $R$ and the field strength $\xi$.

Fig. 10\quad The scaled radial velocity $\mathfrak{f}\left(  \sqrt{2}%
,0,\xi\right)  =$ at $\left(  R=\sqrt{2},\vartheta=0\right)  $ as a function
of $\xi$. This is equivalent to Wilson's electrophoretic factor $f\left(
\xi\right)  $, which is calculated with $\mathbf{v}_{x}$ in cylindrical
coordinates instead of spherical coordinates.

Fig. 11\quad The scaled radial velocity $\mathfrak{f}\left(  1/\sqrt{2}%
,0,\xi\right)  =$ at $\left(  R=1/\sqrt{2},\vartheta=0\right)  $ as a function
of $\xi$. Comparison of this figure with Fig. 10 gives an idea how the
electrophoresis might vary with $R$.\newpage%

%TCIMACRO{\FRAME{ftbpFU}{3.039in}{2.495in}{0pt}{\Qcb{ }}{\Qlb{fig1}%
%}{figure1.jpg}{\special{ language "Scientific Word";  type "GRAPHIC";
%maintain-aspect-ratio TRUE;  display "USEDEF";  valid_file "F";
%width 3.039in;  height 2.495in;  depth 0pt;  original-width 6.9272in;
%original-height 5.6749in;  cropleft "0";  croptop "1";  cropright "1";
%cropbottom "0";
%filename 'New Figures-II/Figure1.jpg';file-properties "XNPEU";}} }%
%BeginExpansion
\begin{figure}
[ptb]
\begin{center}
\includegraphics[
natheight=5.674900in,
natwidth=6.927200in,
height=2.495in,
width=3.039in
]%
{New 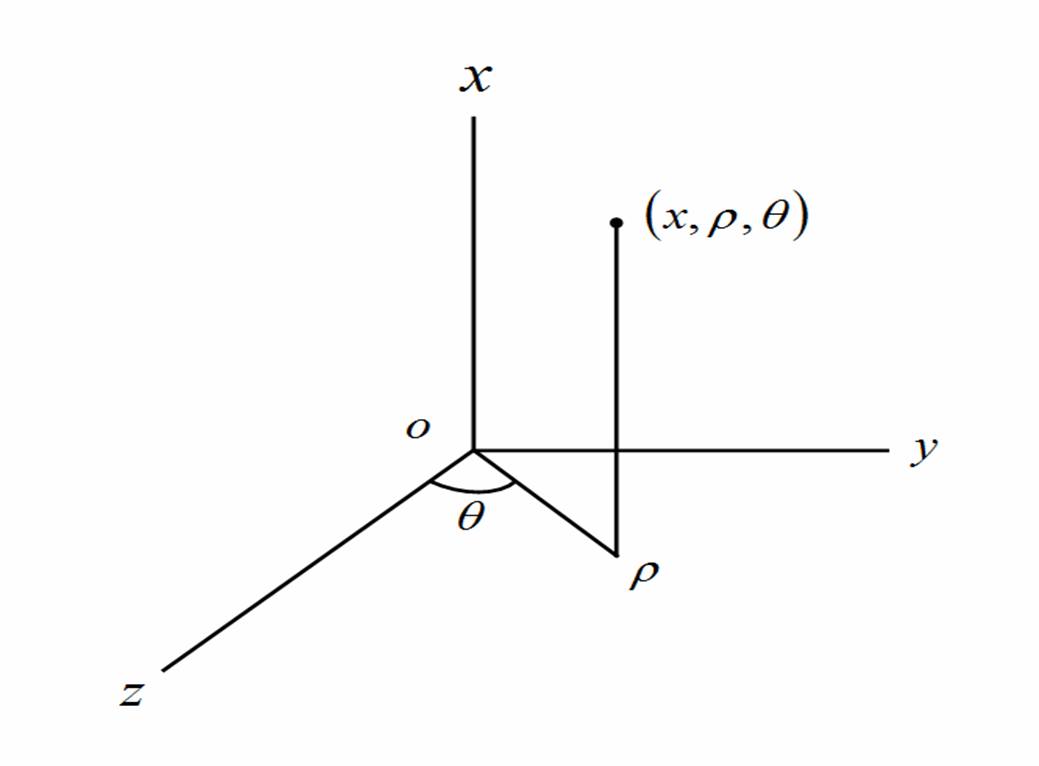}%
\caption{ }%
\label{fig1}%
\end{center}
\end{figure}
%EndExpansion
\newpage%

%TCIMACRO{\FRAME{ftbpFU}{3.039in}{2.6299in}{0pt}{\Qcb{ }}{\Qlb{fig2}%
%}{figure2.jpg}{\special{ language "Scientific Word";  type "GRAPHIC";
%maintain-aspect-ratio TRUE;  display "USEDEF";  valid_file "F";
%width 3.039in;  height 2.6299in;  depth 0pt;  original-width 7.0067in;
%original-height 6.0546in;  cropleft "0";  croptop "1";  cropright "1";
%cropbottom "0";
%filename 'New Figures-II/Figure2.jpg';file-properties "XNPEU";}} }%
%BeginExpansion
\begin{figure}
[ptb]
\begin{center}
\includegraphics[
natheight=6.054600in,
natwidth=7.006700in,
height=2.6299in,
width=3.039in
]%
{New 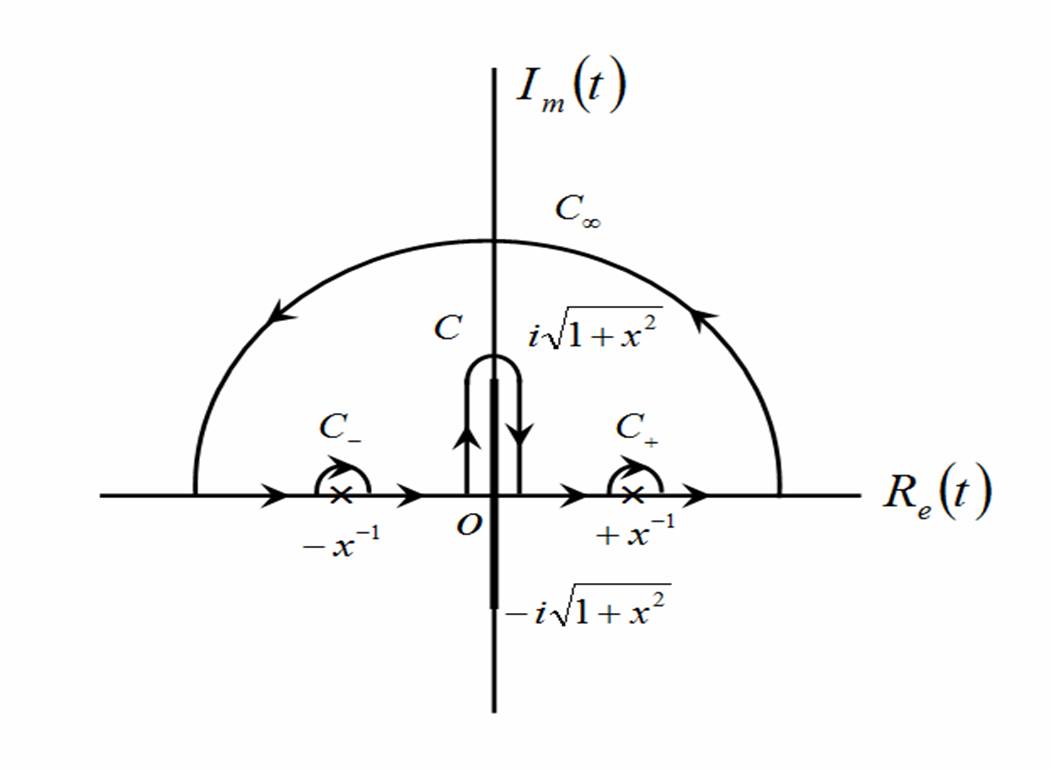}%
\caption{ }%
\label{fig2}%
\end{center}
\end{figure}
%EndExpansion
\newpage%

%TCIMACRO{\FRAME{ftbpFU}{3.039in}{2.6818in}{0pt}{\Qcb{ }}{\Qlb{fig3}%
%}{figure3.jpg}{\special{ language "Scientific Word";  type "GRAPHIC";
%maintain-aspect-ratio TRUE;  display "USEDEF";  valid_file "F";
%width 3.039in;  height 2.6818in;  depth 0pt;  original-width 6.8536in;
%original-height 6.039in;  cropleft "0";  croptop "1";  cropright "1";
%cropbottom "0";
%filename 'New Figures-II/Figure3.jpg';file-properties "XNPEU";}} }%
%BeginExpansion
\begin{figure}
[ptb]
\begin{center}
\includegraphics[
natheight=6.039000in,
natwidth=6.853600in,
height=2.6818in,
width=3.039in
]%
{New 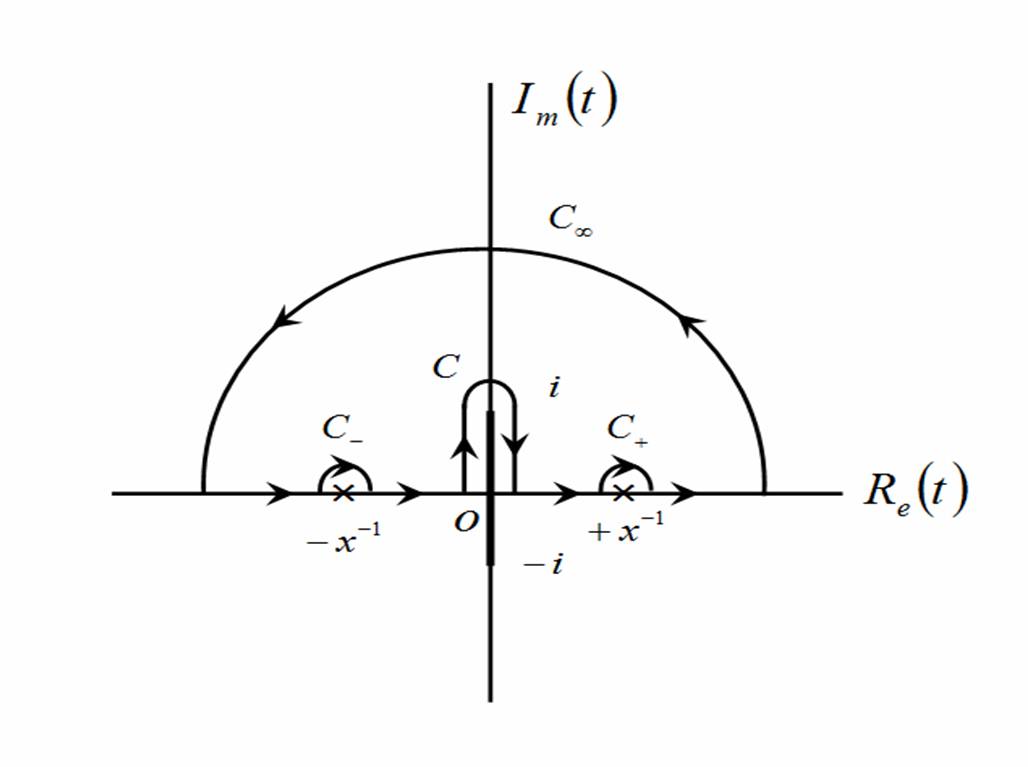}%
\caption{ }%
\label{fig3}%
\end{center}
\end{figure}
%EndExpansion
\newpage%

%TCIMACRO{\FRAME{ftbpFU}{3.0459in}{2.738in}{0pt}{\Qcb{ }}{\Qlb{fig4}%
%}{figure4.jpg}{\special{ language "Scientific Word";  type "GRAPHIC";
%display "USEDEF";  valid_file "F";  width 3.0459in;  height 2.738in;
%depth 0pt;  original-width 6.7732in;  original-height 6.0952in;
%cropleft "0";  croptop "1";  cropright "1";  cropbottom "0";
%filename 'New Figures-II/Figure4.jpg';file-properties "XNPEU";}} }%
%BeginExpansion
\begin{figure}
[ptb]
\begin{center}
\includegraphics[
natheight=6.095200in,
natwidth=6.773200in,
height=2.738in,
width=3.0459in
]%
{New 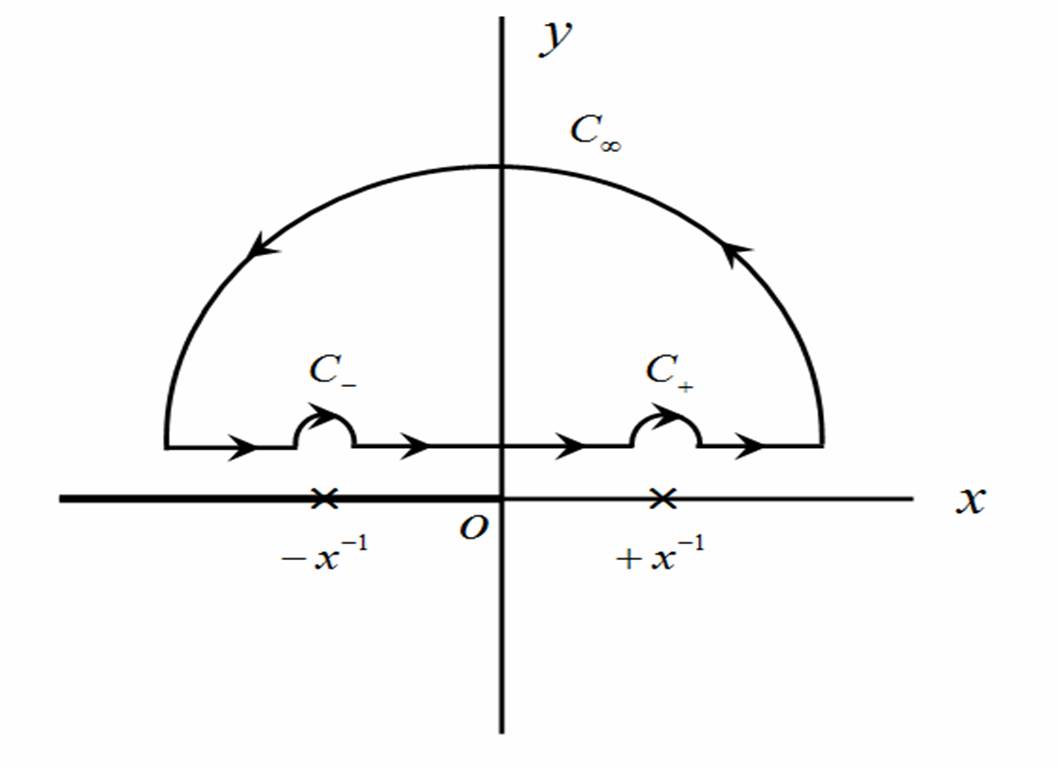}%
\caption{ }%
\label{fig4}%
\end{center}
\end{figure}
%EndExpansion
\newpage%

%TCIMACRO{\FRAME{ftbpFU}{3.039in}{2.2822in}{0pt}{\Qcb{ }}{\Qlb{fig5}%
%}{figure5.tif}{\special{ language "Scientific Word";  type "GRAPHIC";
%maintain-aspect-ratio TRUE;  display "USEDEF";  valid_file "F";
%width 3.039in;  height 2.2822in;  depth 0pt;  original-width 7.9796in;
%original-height 5.9733in;  cropleft "0";  croptop "1";  cropright "1";
%cropbottom "0";
%filename 'New Figures-II/Figure5.tif';file-properties "XNPEU";}} }%
%BeginExpansion
\begin{figure}
[ptb]
\begin{center}
\includegraphics[
natheight=5.973300in,
natwidth=7.979600in,
height=2.2822in,
width=3.039in
]%
{New Figures-II/Figure5.tif}%
\caption{ }%
\label{fig5}%
\end{center}
\end{figure}
%EndExpansion
\newpage%

%TCIMACRO{\FRAME{ftbpFU}{3.039in}{2.2822in}{0pt}{\Qcb{ }}{\Qlb{fig6}%
%}{figure6.tif}{\special{ language "Scientific Word";  type "GRAPHIC";
%maintain-aspect-ratio TRUE;  display "USEDEF";  valid_file "F";
%width 3.039in;  height 2.2822in;  depth 0pt;  original-width 7.9796in;
%original-height 5.9733in;  cropleft "0";  croptop "1";  cropright "1";
%cropbottom "0";
%filename 'New Figures-II/Figure6.tif';file-properties "XNPEU";}} }%
%BeginExpansion
\begin{figure}
[ptb]
\begin{center}
\includegraphics[
natheight=5.973300in,
natwidth=7.979600in,
height=2.2822in,
width=3.039in
]%
{New Figures-II/Figure6.tif}%
\caption{ }%
\label{fig6}%
\end{center}
\end{figure}
%EndExpansion
\newpage%

%TCIMACRO{\FRAME{ftbpFU}{3.039in}{2.2822in}{0pt}{\Qcb{ }}{\Qlb{fig7}%
%}{figure7.tif}{\special{ language "Scientific Word";  type "GRAPHIC";
%maintain-aspect-ratio TRUE;  display "USEDEF";  valid_file "F";
%width 3.039in;  height 2.2822in;  depth 0pt;  original-width 7.9796in;
%original-height 5.9733in;  cropleft "0";  croptop "1";  cropright "1";
%cropbottom "0";
%filename 'New Figures-II/Figure7.tif';file-properties "XNPEU";}} }%
%BeginExpansion
\begin{figure}
[ptb]
\begin{center}
\includegraphics[
natheight=5.973300in,
natwidth=7.979600in,
height=2.2822in,
width=3.039in
]%
{New Figures-II/Figure7.tif}%
\caption{ }%
\label{fig7}%
\end{center}
\end{figure}
%EndExpansion
\newpage%

%TCIMACRO{\FRAME{ftbpFU}{3.039in}{2.2822in}{0pt}{\Qcb{ }}{\Qlb{fig8}%
%}{figure8.tif}{\special{ language "Scientific Word";  type "GRAPHIC";
%maintain-aspect-ratio TRUE;  display "USEDEF";  valid_file "F";
%width 3.039in;  height 2.2822in;  depth 0pt;  original-width 7.9796in;
%original-height 5.9733in;  cropleft "0";  croptop "1";  cropright "1";
%cropbottom "0";
%filename 'New Figures-II/Figure8.tif';file-properties "XNPEU";}} }%
%BeginExpansion
\begin{figure}
[ptb]
\begin{center}
\includegraphics[
natheight=5.973300in,
natwidth=7.979600in,
height=2.2822in,
width=3.039in
]%
{New Figures-II/Figure8.tif}%
\caption{ }%
\label{fig8}%
\end{center}
\end{figure}
%EndExpansion
\newpage%

%TCIMACRO{\FRAME{ftbpFU}{3.039in}{2.2805in}{0pt}{\Qcb{ }}{\Qlb{fig9}%
%}{figure9.tif}{\special{ language "Scientific Word";  type "GRAPHIC";
%maintain-aspect-ratio TRUE;  display "USEDEF";  valid_file "F";
%width 3.039in;  height 2.2805in;  depth 0pt;  original-width 7.9874in;
%original-height 5.9733in;  cropleft "0";  croptop "1";  cropright "1";
%cropbottom "0";
%filename 'New Figures-II/Figure9.tif';file-properties "XNPEU";}} }%
%BeginExpansion
\begin{figure}
[ptb]
\begin{center}
\includegraphics[
natheight=5.973300in,
natwidth=7.987400in,
height=2.2805in,
width=3.039in
]%
{New Figures-II/Figure9.tif}%
\caption{ }%
\label{fig9}%
\end{center}
\end{figure}
%EndExpansion
\newpage%

%TCIMACRO{\FRAME{ftbpFU}{3.039in}{2.2822in}{0pt}{\Qcb{ }}{\Qlb{fig10}%
%}{figure10.tif}{\special{ language "Scientific Word";  type "GRAPHIC";
%maintain-aspect-ratio TRUE;  display "USEDEF";  valid_file "F";
%width 3.039in;  height 2.2822in;  depth 0pt;  original-width 7.9796in;
%original-height 5.9733in;  cropleft "0";  croptop "1";  cropright "1";
%cropbottom "0";
%filename 'New Figures-II/Figure10.tif';file-properties "XNPEU";}} }%
%BeginExpansion
\begin{figure}
[ptb]
\begin{center}
\includegraphics[
natheight=5.973300in,
natwidth=7.979600in,
height=2.2822in,
width=3.039in
]%
{New Figures-II/Figure10.tif}%
\caption{ }%
\label{fig10}%
\end{center}
\end{figure}
%EndExpansion
\newpage%

%TCIMACRO{\FRAME{ftbpFU}{3.039in}{2.2822in}{0pt}{\Qcb{ }}{\Qlb{fig11}%
%}{figure11.tif}{\special{ language "Scientific Word";  type "GRAPHIC";
%maintain-aspect-ratio TRUE;  display "USEDEF";  valid_file "F";
%width 3.039in;  height 2.2822in;  depth 0pt;  original-width 7.9796in;
%original-height 5.9733in;  cropleft "0";  croptop "1";  cropright "1";
%cropbottom "0";
%filename 'New Figures-II/Figure11.tif';file-properties "XNPEU";}} }%
%BeginExpansion
\begin{figure}
[ptb]
\begin{center}
\includegraphics[
natheight=5.973300in,
natwidth=7.979600in,
height=2.2822in,
width=3.039in
]%
{New Figures-II/Figure11.tif}%
\caption{ }%
\label{fig11}%
\end{center}
\end{figure}
%EndExpansion

\end{document}